%% file: main.tex
\pdfoutput=1

\newif\ifAnon\Anonfalse
\newif\ifDraft\Draftfalse
\newif\ifSubmission\Submissiontrue

\documentclass[conference]{IEEEtran}
\IEEEoverridecommandlockouts
\usepackage[hyphens]{url}
\PassOptionsToPackage{bookmarks,hidelinks}{hyperref}
\usepackage{hyperref}
\hypersetup{breaklinks=true}
\usepackage{epsfig,endnotes}
\usepackage{tugraz_defaults}
\usepackage{csquotes}
\usepackage{flushend}
\usepackage{subcaption}

\DeclareMathAlphabet\mathbfcal{OMS}{cmsy}{b}{n}
\newcommand{\COne}{$\mathbfcal{C}\mathbf{1}$\xspace}
\newcommand{\CTwo}{$\mathbfcal{C}\mathbf{2}$\xspace}
\newcommand{\CThree}{$\mathbfcal{C}\mathbf{3}$\xspace}
\newcommand{\CFour}{$\mathbfcal{C}\mathbf{4}$\xspace}

\newcommand{\POne}{$\mathbfcal{P}\mathbf{1}$\xspace}
\newcommand{\PTwo}{$\mathbfcal{P}\mathbf{2}$\xspace}

\newcommand{\paragrabf}[1]{\textbf{#1}.\ }

\newcommand{\srcfile}{\raisebox{-0.2em}{\includegraphics[height=1em]{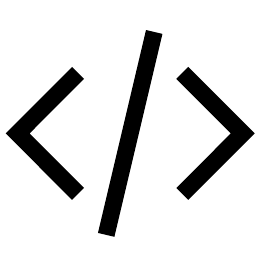}}}
\newcommand{\binfile}{\raisebox{-0.2em}{\includegraphics[height=1em]{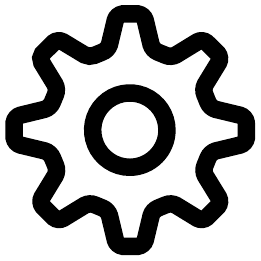}}}

\renewcommand{\paragraph}[1]{\vspace{0.0cm}\noindent\textbf{#1}\ }

\usepackage{balance}

\makeatletter
\newcommand{\linebreakand}{%
  \end{@IEEEauthorhalign}
  \hfill\mbox{}\par
  \mbox{}\hfill\begin{@IEEEauthorhalign}
}
\makeatother

\begin{document}
\title{Automating Seccomp Filter Generation for Linux Applications}
\date{}

\ifAnon
\author{}
\else
\author{
  \IEEEauthorblockN{
    Claudio Canella\IEEEauthorrefmark{1},
    Mario Werner\IEEEauthorrefmark{1},
    Daniel Gruss\IEEEauthorrefmark{1},
    Michael Schwarz\IEEEauthorrefmark{3},
  }
  \bigskip
  \IEEEauthorblockA{
      \IEEEauthorrefmark{1}Graz University of Technology
      \qquad
      \IEEEauthorrefmark{3}CISPA Helmholtz Center for Information Security
  }
}
\fi

\newcommand{\ToolName}{Chestnut\xspace}
\newcommand{\BinName}{Binalyzer\xspace}
\newcommand{\CompilerName}{Sourcalyzer\xspace}
\newcommand{\DynamicName}{Finalyzer\xspace}
\newcommand{\RewriteTool}{ChestnutPatcher\xspace}
\newcommand{\SandboxTool}{ChestnutGenerator\xspace}
\newcommand{\LibraryName}{libchestnut\xspace}
\newcommand{\syscall}{syscall\xspace}
\newcommand{\Syscall}{Syscall\xspace}
\newcommand{\syscalls}{syscalls\xspace}
\newcommand{\Syscalls}{Syscalls\xspace}

\maketitle
\thispagestyle{plain}
\pagestyle{plain}

\begin{abstract}
Software vulnerabilities in applications undermine the security of applications.
By blocking unused functionality, the impact of potential exploits can be reduced.
While seccomp provides a solution for filtering \syscalls, it requires manual implementation of filter rules for each individual application.
Recent work has investigated automated approaches for detecting and installing the necessary filter rules.
However, as we show, these approaches make assumptions that are not necessary or require overly time-consuming analysis.

In this paper, we propose \ToolName, an automated approach for generating strict \syscall filters for Linux userspace applications with lower requirements and limitations.
\ToolName comprises two phases, with the first phase consisting of two static components, \ie a compiler and a binary analyzer, that extract the used \syscalls during compilation or in an analysis of the binary.
The compiler-based approach of \ToolName is up to factor \SIx{73} faster than previous approaches without affecting the accuracy adversely.
On the binary analysis level, we demonstrate that the requirement of position-independent binaries of related work is not needed, enlarging the set of applications for which \ToolName is usable.
In an optional second phase, \ToolName provides a dynamic refinement tool that allows restricting the set of allowed \syscalls further.
We demonstrate that \ToolName on average blocks \SIx{302} \syscalls (\SI{86.5}{\percent}) via the compiler and \SIx{288} (\SI{82.5}{\percent}) using the binary-level analysis on a set of 18 widely used applications.
We found that \ToolName blocks the dangerous \texttt{exec} \syscall in \SI{50}{\percent} and \SI{77.7}{\percent} of the tested applications using the compiler- and binary-based approach, respectively.
For the tested applications, \ToolName prevents exploitation of more than \SI{62}{\percent} of the \SIx{175} CVEs that target the kernel via \syscalls.
Finally, we perform a 6 month long-term study of a sandboxed Nginx server.
\end{abstract}

\section{Introduction} \label{sec:intro}
The complexity of applications is steadily growing~\cite{Holzmann2015,Quach2017}, and with that, also the number of vulnerabilities found in applications~\cite{Miller2019Bluehat}.
A consequence is that the attack surface for exploits is also growing.
Especially in applications written in memory unsafe languages such as C, bugs often lead to memory safety violations that potentially enable exploits~\cite{Szekeres2013sok}.
While state-of-the-art defensive-programming techniques and countermeasures reduce the number of vulnerabilities, there is still a remaining risk that an attacker can exploit a vulnerability in an application.
Especially for privileged applications such as \texttt{setuid} binaries, this can, in the worst case, mean that an attacker can take over the entire system.

The remaining exploitation risk can be addressed by reducing the post-exploitation impact (\cf principle of least privilege).
With available resources and interfaces limited to those strictly required by the application, a successful exploit cannot use arbitrary other functionality~\cite{Lin2018container}.
Especially blocking dangerous \syscalls and \syscall parameters that are not required by many applications, \eg the \texttt{exec} \syscall to execute a new program, reduces an attacker's possibilities in the post-exploitation phase.
Application sandboxing limits the resources available to an application~\cite{Prevelakis2001sandboxing,Goldberg1996secure} and, ideally, untrusted and potentially malicious, or benign but compromised applications cannot escape the sandbox.

On Linux, seccomp~\cite{Edge2015seccomp} and the extended seccomp-bpf can be used by applications to restrict the \syscall interface.
Seccomp-bpf~\cite{Edge2015seccomp} supports filter rules via developer-defined Berkeley Packet Filters~\cite{McCanne1993bsd}.
Each \syscall can either be entirely blocked or specific arguments for it.
However, the correct usage of seccomp-bpf requires the developer to know which \syscalls are used by the application and the included libraries.
As this is a considerable effort, seccomp is mainly used in applications that implement isolation mechanisms, \eg sandboxes~\cite{Google2019sandbox2,firejail2016firejail}.
Given its complexity, it is rarely used in other applications.

Recent works proposed two methods to automatically generate such seccomp filters~\cite{Ghavamnia2020,DeMarinis2020}.
The first approach utilizes the compiler and various external tools to derive the filters during compilation~\cite{Ghavamnia2020}.
To minimize the set of \syscalls, the approach relies on sophisticated points-to analysis~\cite{Andersen1994points} to generate a call graph of reachable functions and \syscalls.
The second approach relies on binary analysis to determine the \syscalls an existing binary intends to use~\cite{DeMarinis2020}.
While these are first solutions to the problem of automating filter generation, both come with clear limitations.
For instance, the first approach does not scale with the program size due to the points-to analysis~\cite{Andersen1994points,Hind2001pointer}.
In practice, the overheads can be prohibitively large as they would require a massive upscaling of development and build server resources.
The second approach comes with a strong requirement that the application is compiled as a position-independent code (PIC) binary (PIE)~\cite{DeMarinis2020}.
While PIE is the default on recent Ubuntu distributions for C and C++ compiled programs, static C and C++ binaries are by default not compiled as PIE.
Other compiled binaries are often not PIE either, \eg `golang' binaries such as the popular git server Gogs, which are not supported by these previous works.
Both limitations reduce the set of applications that can be protected with these solutions substantially.

In this paper, we present a novel approach that overcomes the limitations of previous ones and automatically generates strict seccomp filters for native Linux userspace applications.
We show that our approach is a significant improvement over the compiler-based approach by Ghavamnia~\etal\cite{Ghavamnia2020} as it does not require a sophisticated points-to analysis to generate filter rules.
Instead, a faster \textit{has address taken} approach can be used that achieves the same accuracy but at a fraction of the performance impact on compilation time.
Second, we demonstrate an alternative implementation to the one provided by DeMarinis~\etal\cite{DeMarinis2020}.
In this approach, we demonstrate that the requirement of a PIC binary is not necessary, significantly extending the set of applications to which it can be applied.
We implement our method in a proof-of-concept tool, \ToolName.\footnote{The prototype and several demo videos can be found in our anonymous GitHub repository \url{https://github.com/chestnut-sandbox/Chestnut}.}
We also advance the state of the art in evaluations of automatic syscall filtering, with a first long-term case study and coverage metrics to confirm our approach's validity.

\ToolName uses a two-phase process.
A static first phase \POne consisting of two static components (\textbf{\CompilerName} and \textbf{\BinName}), and an optional dynamic second phase \PTwo (\textbf{\DynamicName}).
Based on static analysis, \ToolName first identifies the set of unused \syscalls without running the application in \POne and dynamically refines this set in \PTwo to reduce the inherent limitations of the static analysis in \POne.

For \textbf{\CompilerName}, we extend the LLVM framework to detect the \syscalls used by the application already during compile- and link-time.
The \syscall information for each shared library is either extracted using the compiler or using \BinName. 
\textbf{\BinName} can be used for applications and libraries which are either not compatible with LLVM or where the source code is not available.
We rely on capstone~\cite{Quynh2014capstone} to disassemble applications and to locate \syscalls.
Using symbolic backward execution~\cite{Ma2011directed} from the \texttt{syscall} instruction, we infer the \syscall number used in the identified \syscall.
Additionally, we use the control-flow graph recovery functionality of angr~\cite{Wang2017angr} to map exported functions to identified \syscalls.
Exactly as in previous work, an inherent limitation of static approaches is that they can miss \syscalls in rare cases if control-flow cannot be inferred correctly.
However, we observe that more frequently, the set of used \syscalls is overapproximated.
To refine the number of allowed \syscalls, we provide a complementary optional dynamic approach in the second phase of \ToolName.
In this second phase, \textbf{\DynamicName} traces all \syscalls of the application and then refines the allowlist to further restrict or relax the seccomp filters.

To demonstrate our approach's feasibility, we evaluate various real-world client applications, such as git and busybox, database applications such as redis and sqlite3, and Nginx as a server application.
We show that \ToolName does not impair their functionality while it significantly reduces the attack surface.
On average, \ToolName blocks \SIx{295} \syscalls (\SI{84.5}{\percent}) on Linux kernel 5.0.
In the \SIx{18} real-world binaries we evaluated, \ToolName blocked the \texttt{exec} \syscall for \SI{50}{\percent} of the applications using \CompilerName and in \SI{77.7}{\percent} using \BinName.
We prevent the \texttt{mprotect} \syscall in \SI{61.1}{\percent} of the tested applications using \CompilerName.
Furthermore, we evaluate our approach with existing real-world exploits, showing that \ToolName prevents exploitation of around \SI{64}{\percent} and \SI{62}{\percent} of CVEs using \CompilerName and \BinName, respectively.
We also compare our approaches with related previous work~\cite{Ghavamnia2020,DeMarinis2020}.
We show that we can achieve similar results in terms of effectiveness, measured in the practically mitigated CVEs, to the compiler-based approach by Ghavamnia~\etal\cite{Ghavamnia2020} but improving the performance by up to factor \SIx{73}.
We are the first to show that binary-based approaches can also be applied to non-PIC binaries by demonstrating that \BinName runs successfully on non-PIC binaries.
We evaluate the functional correctness of \CompilerName in functional tests as well as a 6-month long-term case study:
During 6 months of use of a \CompilerName-protected Nginx production server, we did not observe a single crash.
Furthermore, we are the first to evaluate how tight automatically generated filter rules actually are.
We evaluate the functional correctness and the tightness of the filters by executing the available test suites.
We substantiate the validity of these experiments by measuring the code coverage of the respective test suites.

Filters generated automatically with a tool might not always be as strict as theoretically possible.
However, there is no time investment required from the developer, making it a very inexpensive defense in depth.
More importantly, \ToolName can be applied to and improve the security of existing and widely-used technology, \ie seccomp, making \syscall filtering available to commodity applications.
The only runtime overhead introduced is the small overhead of using seccomp, similar as containers already do today.

To summarize, we make the following contributions:
\begin{compactenum}
  \item We present a novel compiler-based approach for automatic \syscall-filter generation without manual interaction that is up to factor \SIx{73} faster than previous work.
  \item We present a method to refine the number of allowed \syscalls based on dynamic tracing.
  \item We demonstrate that \ToolName prevents the exploitation of more than \SI{63}{\percent} of the \SIx{175} CVEs in the Linux kernel exploitable via \syscalls.
  \item We show that requirements of previous approaches are not necessary, thus enabling a significantly faster approach that is also applicable to a wider range of applications.
  \item We perform a 6 month long-term study using Nginx to demonstrate the functional correctness of our approach where we did not observe a single crash.
\end{compactenum}

\textbf{Outline.}
\cref{sec:background} provides background.
In \cref{sec:design}, we discuss the threat model, challenges, and design of \ToolName.
In \cref{sec:static-sandboxing}, we detail our compiler-based approach and the extraction from existing binaries.
\cref{sec:dynamic-sandboxing} discusses our dynamic refinement approach.
We evaluate \ToolName in \cref{sec:evaluation}.
We discuss related work, limitations, and future work in \cref{sec:discussion}.
We conclude in \cref{sec:conclusion}.

\section{Background}\label{sec:background}

\subsection{Sandboxing}
Sandboxing is a security mechanism that intends to constrain software within a tightly controlled environment by restricting the available resources to a required minimum~\cite{Prevelakis2001sandboxing,Goldberg1996secure}.
Hence, the damage in case of exploitation is limited.
These restrictions may encompass the ability to access the network, limit the amount of storage, file descriptors, or inhibit the application from issuing specific \syscalls.
By now, different forms of sandboxing have been adopted by many browser vendors to secure their products~\cite{Sylvain2015,Firefox2019sandbox,Narayan2020retrofitting,Reis2009,Reis2019SiteIsolation,Firefox2019fission}.

\subsection{Linux Seccomp}
To facilitate operations that require higher privileges or direct hardware access, the kernel provides \syscalls to every userspace application.
As with other interfaces, they also contain bugs that can lead to privilege escalation~\cite{Kemerlis2014,Kemerlis2015,Kemerlis2012}.
Hence, platform security profits from limiting the amount of \syscalls that an application can perform.
With Secure Computing (seccomp)~\cite{Edge2015seccomp}, Linux provides a filter that allows a userspace program to specify the \syscalls it performs over its lifetime.
The kernel then blocks the remaining \syscalls for the sandboxed application that might originate from an application being exploited.
As seccomp filters do not dereference pointers, so-called time-of-check time-of-use attacks~\cite{McPhee1972integrity} common in \syscall interposition frameworks are not possible. %
Examples of applications that rely on seccomp are Chromium~\cite{ChromiumSeccomp}, Firefox~\cite{Mozilla2016seccomp}, and the zygote process in Android systems~\cite{Android2017seccomp}.

\subsection{Memory Safety}

Memory safety is an essential concept in computer security, and its violation can lead to exploitation.
One way to exploit a program is to corrupt its memory and to divert control flow to a previously injected code sequence.
This code sequence, \ie the payload, is called \textit{shellcode} and is commonly written in machine code.
These types of attacks are commonly referred to as control-flow hijack attacks~\cite{Szekeres2013sok}.

Nergal~\cite{Nergal2001ret2libc} and Shacham~\cite{Shacham2007} describe ROP attacks, which allow an attacker to chain existing code gadgets within an application together to perform malicious tasks.
Each such gadget is a sequence of instructions that end with a return instruction.
ROP attacks are hard to defend as all the information is already present within the application, \ie an attacker does not need to inject code.
While ROP attacks overwrite saved return addresses, similar attacks exist that overwrite other pointers~\cite{Checkoway2010JOP,Bletsch2011JOP,Lan2015loop,Carlini2014COP,Goktas2014COP,Schuster2015COOP} or signal handlers~\cite{Bosman2014SROP}.

\subsection{Executable and Linkable Format}
On Unix-based systems, the Executable and Linkable Format (ELF)~\cite{Bovet2013elf,Drysdale2015elf} is the standard file format for shared libraries and executable files.
One advantage is that it is highly flexible and extensible.
An ELF file consists of an ELF header that is followed by data.
The data itself can consist of a program and a section header table describing segments and sections, respectively.
Segments contain information that is necessary for the run-time execution of the ELF binary, while sections contain information relevant for linking and relocating.

\paragraph{Dynamic Linking.}
The dynamic linker is responsible for loading and linking shared libraries needed by an executable during runtime~\cite{Drysdale2015elf}.
For that, the dynamic linker copies the shared library's content into memory and ensures its functionality, \eg filling jump tables and relocating pointers.
On Unix-like systems, the dynamic linker is selected during link time and is embedded into the ELF file. %

\section{Design of \ToolName}\label{sec:design}

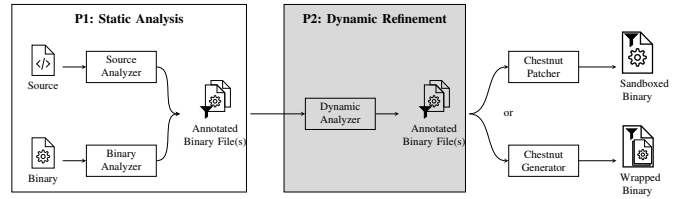
\begin{figure}
  \centering
  \resizebox{\hsize}{!}{
    \input{images/overview.tikz}
  }
 \caption{The components of \ToolName and their interaction.
  In \POne, source files can be analyzed statically with our LLVM-based analyzer, binaries with our binary analyzer.
  If necessary, dynamic analysis in the optional \PTwo refines the previous filters.
  After both phases, the binary can either be rewritten to block unused \syscalls or a tailored sandbox can be generated that allows only the \syscalls used in the binary. }
 \label{fig:overview}
\end{figure}

In this section, we introduce the threat model used throughout the paper.
We outline a set of challenges required to solve automatic filter generation and discuss the high-level idea of \ToolName.
\Cref{fig:overview} illustrates the main components of \ToolName, \eg the compiler modification \textit{\CompilerName}, the binary analyzer \textit{\BinName}, and the dynamic refinement tool \textit{\DynamicName}.
Furthermore, we discuss how all three components can be combined to further enhance the capabilities of \ToolName.

\subsection{Threat Model and Idea of \ToolName}\label{sec:threat-model}
\ToolName supports Linux applications available as either C source code or as a binary.
In the latter case, it is not limited to PIC binaries as previous work has discussed~\cite{DeMarinis2020}.
These applications can range from server applications to applications executing potentially malicious code that is not controlled by the user, such as browsers, office applications~\cite{Mueller2020,Infuer2019}, and pdf readers~\cite{Esparza2012,Esparza2014}.
Another example where \ToolName can be used to restrict the \syscall interface are messenger applications as several attacks have been shown to fully compromise a system~\cite{Silvanovich2020,Gross2020}.
We assume correct usage of \ToolName in one of its variants (\cf \Cref{fig:overview}).
\ToolName assumes that the application itself is not malicious but potentially vulnerable to exploitation, \eg due to a memory-safety violation, enabling an attacker to gain arbitrary code execution within a vulnerable application.
We assume that the post-exploitation step targets the trusted system and requires \syscalls, \eg to gain kernel privileges.
\ToolName prevents the application from using \syscalls it usually would not use, averting harm from the rest of the system.
\Syscalls provided for file operations can potentially be exploited by an attacker to modify configuration files.
Argument-level API specilization~\cite{mishra2018shredder} can be used to protect against such attacks. 
In line with related work~\cite{Ghavamnia2020}, \ToolName currently cannot protect against such attacks if the application requires these \syscalls, and the file permissions are set incorrectly.
Augmenting \ToolName with argument-level API specialization is left for future work.
\ToolName is orthogonal to other defenses such as CFI, ASLR, NX, or canary-based protections and enhances the security in case these other mitigations have been circumvented.
Side-channel and fault attacks~\cite{Kim2014,Yarom2014Flush,Kocher2019,Lipp2018meltdown,VanSchaik2019RIDL,Schwarz2019ZL,Canella2019A} are out of scope.

\subsection{Challenges}
Automatic filter generation using a static approach requires solving the following four challenges:

\paragraph{\COne: Identifying \Syscall Numbers for each \Syscall.}
To automatically block unused \syscalls, it is necessary to identify the \syscalls used by the application.
The \syscall itself is usually a single instruction, \eg \texttt{syscall} (x86\_64) or \texttt{svc \#0} (AArch64).
The actual \syscall is specified as a number in a CPU register, \eg \texttt{rax} (x86\_64) or \texttt{x8} (AArch64)~\cite{Drysdale2014syscall}.
Hence, the first challenge is identifying the individual \syscall number that a specific \syscall uses.
\Syscalls can appear in many different forms within a binary, \eg inline assembly, assembly file, or issued with the libc \textit{syscall} wrapper function.
Moreover, \syscalls might not be called directly, but via a call chain through various libraries.
For example, an application calls \textit{foo()} in \textit{libfoo.so} that calls \textit{open()} in \textit{libc.so}, which finally calls \textit{syscall()} in \textit{libc.so} issuing the \syscall.
For our approach, we have to detect all \syscalls, regardless of how they are called.
Extracting the \syscall number is the only architecture-dependent part of \ToolName.

\paragraph{\CTwo: Reconstruct Call Sites of \Syscalls.}
By solving challenge \COne, we know which \syscalls can be potentially called by the target application.
Unfortunately, including all detected \syscalls of the binary and the used libraries does not suffice.
Most binaries link against libc, which provides an implementation of almost all \syscalls.
Hence, the generated filters would be too permissive as they would basically allow all \syscalls.
We have to analyze the reachability of the identified \syscalls by constructing a call graph for every binary.
This call graph contains the information obtained from \COne for each function and the information about edges between them.

\paragraph{\CThree: Generate Set of \Syscalls.}
To generate the final set of \syscalls for our application, the information from \COne and \CTwo has to be combined for the application and its libraries.
By combining the call graph obtained in \CTwo with the information which functions are used in the application and libraries, we create a set of functions potentially called by the application.
In combination with the call graph (\CTwo) and \syscall numbers (\COne), this set provides the information about all the \syscalls that the application can execute.

\paragraph{\CFour: Install Filters.}
We rely on seccomp to apply the \syscall filters as seccomp is natively supported on Linux.
To install the filters, we provide a library (\LibraryName) that has to be added to the application.
This library uses the allowlist (\CThree), generates the seccomp rules, and installs the resulting filters before the actual application starts at the main entry point.

Once all these challenges have been solved, we automatically obtain an application that can only execute the required \syscalls.
We solve the challenges in detail in \cref{sec:static-sandboxing,sec:dynamic-sandboxing}.

\subsection{High-Level Idea} \label{sec:high-level}
This section briefly discusses the three components \ToolName provides in the two phases (\POne and \PTwo) and how they solve the challenges.
\cref{sec:static-sandboxing,sec:dynamic-sandboxing} provide more detail on the implementation of each component.

\paragraph{\CompilerName.}
\ToolName contains \CompilerName, a compiler-based component for static analysis of the application source code.
Based on LLVM, \CompilerName is a compiler pass that extracts all \syscalls identified during compilation.

This proves to be a practical approach for statically linked binaries.
For such applications, and given that libraries are compiled with \CompilerName, the compiler and linker are aware of the entire codebase and can thus identify every \syscall instruction of the final binary.
Unfortunately, just extracting the numbers and installing a seccomp filter for all found \syscalls is not enough, as this would lead to almost all \syscalls being allowed.
The reason is that the C standard library implements almost all \syscalls.
By linking against it, our generated filters would allow almost all \syscalls, which renders the filters ineffective.
Hence, we need to determine further which \syscalls are actually used by the application by analyzing the control-flow graph to solve challenges \CTwo and \CThree.
While comparable work~\cite{Ghavamnia2020} needs to perform the same task, we demonstrate a solution that is up to factor \SIx{73} faster.
We discuss this in \cref{sec:static-compiler}.

\paragraph{\BinName.}
The compiler-based approach's limitation is that it requires the source code of the application and all used libraries.
\BinName has the same goal as \CompilerName but works directly on the binary level.
With this, our approach is also applicable to programs where the source code is not available or where the source code is not compatible with LLVM, retrofitting the approach to binaries.
In contrast to previous work~\cite{DeMarinis2020}, \BinName is also not restricted to PIC binaries.

The idea of this binary-level analyzer is to scan binaries and libraries for \textit{syscall} instructions and then use symbolic backward execution~\cite{Ma2011directed} from these locations to infer the respective \syscall number, again solving challenge \COne.
Similar to the compiler-based approach, the basic binary-level approach also suffers from overapproximation.
To reduce overapproximation, \BinName leverages control-flow-graph analysis of all dependencies to map exported functions to the identified \syscall numbers (\CTwo).
Finally, based on the required symbols of the binary and the libraries and the \syscall-to-function mapping, \BinName infers a set of \syscalls reachable by the application (\CThree).
Note that \BinName can also be applied to stripped binaries as all the required information is still included for dynamic linking, \ie the list of exported functions to which we add \syscalls.
We detail this in \cref{sec:binary-extraction}.

\paragraph{\DynamicName.}
To work around the limitations of static analysis, we propose \DynamicName, an approach based on dynamic \syscall tracing. %
\DynamicName is solely intended to refine filters identified by our static approaches in a developer-controlled, benign environment. 
In this optional phase, \DynamicName removes or adds additional filters that cannot be identified statically.

The dynamic nature of \DynamicName allows us to simplify challenges \COne to \CFour by inspecting \syscalls just-in-time.
\DynamicName extracts the \syscall number during runtime (\COne) by intercepting all \syscalls for the target application.
By intercepting the \syscall, it is inherent that the \syscall is reachable (\CTwo).
These dynamically collected \syscalls can then be checked against the statically identified \syscalls.
In this step, missed \syscalls can be added to refine the installed filter list (\CFour).
We discuss this process in more detail in \cref{sec:dynamic-sandboxing}.

\paragraph{Combining Components.}
\ToolName is designed in a way that allows combining all three components, as shown in \cref{fig:overview}.
For instance, \DynamicName is intended to be used as an optional step after the static components if they cannot infer the used \syscalls due to the static analysis's limitations.
An instance where this is necessary is when an application dynamically starts other applications.
The child process inherits the parents' \syscall filters, which cannot be relaxed anymore.
By combining the static approaches with \DynamicName, the \syscalls of the child process can be identified and added to the application's allowlist.
\CompilerName can also be used in combination with \BinName, \eg if the source is available for the application but not for a used library.

\paragraph{Applying \Syscall Filters.}
The output of each component is a file containing the \syscalls the application can call. 
For \CompilerName, the \syscall filters can be directly compiled into the target application.
However, if this is either not desirable or possible, \eg because only the binary is available, we provide two tools to apply the \syscall filters (\cf \cref{fig:overview}).
\SandboxTool creates a sandbox tailored to the target application.
Alternatively, \RewriteTool directly patches the target application to include the \syscall filters and \LibraryName.

\section{Static Filter Extraction}\label{sec:static-sandboxing}
In this section, we present the two static approaches of \POne to automatically generate \syscall filters.
We highlight the necessary steps for solving the outlined challenges in a fast and efficient way in both a compiler and a binary-based approach in more detail.

\subsection{Compiler-Based Approach}\label{sec:static-compiler}
\CompilerName utilizes the LLVM compiler framework~\cite{Lattner2004} to extract \syscalls from source code. 
It uses module passes (\ie one analysis and one transformation pass) that operate on the LLVM intermediate representation (IR).
Additionally, LLVM's linker lld is extended to combine the extracted information from multiple translation units. %
We use an unmodified compiler-rt and musl libc. 
Hence, using \ToolName with the \CompilerName approach is as simple as compiling and linking an application with our extended toolchain.

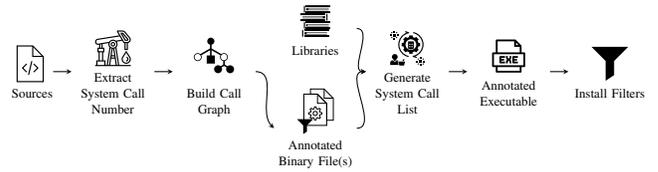
\begin{figure}
  \centering
  \resizebox{\hsize}{!}{
    \input{images/compiler-flow.tikz}
  }
  \caption{The different steps of \CompilerName, which starts with the source and ends with a fully sandboxed application.}
  \label{fig:syscall-extraction}
\end{figure}

\paragrabf{\COne: Identify System Call Numbers}
To invoke a \syscall, x86\_64 provides the \textit{syscall} and AArch64 the \textit{svc \#0} instruction.
The extraction of the \syscall number is the only architecture-dependent part of \ToolName.
Given the \syscall number, the rest of the approach is the same for all architectures supported by LLVM. %
\Syscalls are typically abstracted by the standard C library via the \textit{syscall()} function, \ie \textit{read} invokes the \textit{syscall()} function with the \textit{SYS\_read} \syscall number.
musl additionally provides the function \textit{\_\_syscall\_cp()} that adds a cancellation point to a \syscall.
To detect all invocations of \syscalls, we need to detect all three cases, \ie inline assembly, the \textit{syscall} function, and the \textit{\_\_syscall\_cp()} function.

The LLVM analysis pass iterates over all functions within a translation unit.
For each function, we iterate over every LLVM IR instruction to check whether it is a call site. %
If it is, we check whether it is an inline \syscall assembly statement or a call to either one of the \textit{syscall} or \textit{\_\_syscall\_cp} functions.
In all three cases, we extract the first argument as it is the number of the requested \syscall.
Note that, due to the way we traverse the IR, we also know precisely what function performs the respective \syscall.

Our proof-of-concept implementation currently does not parse assembly files as they are treated differently by LLVM than normal source files.
Hence, if a \syscall is implemented in one, \eg \textit{clone}, we cannot detect it, but a full implementation can handle this case.

\paragrabf{\CTwo: Reconstruct \Syscall Call Sites}
\CompilerName uses the \syscall numbers extracted in \COne as a starting point for further analyzing which \syscalls are used based on the call graph. 
The main challenge in this regard is extracting a reasonably precise call graph without strongly degrading usability due to huge performance overheads or by requiring changes to the common compilation model (\eg by demanding link-time optimization (LTO)).
In particular, restricting indirect function call sites to a set of possible call targets is a necessary, but typically quite expensive, task that commonly relies on inter-procedural pointer analysis (\eg Andersen~\cite{Andersen1994points}, Steensgaard~\cite{Steensgaard1996points}).
This type of analysis requires access to the whole program and often does not scale efficiently to larger program sizes~\cite{Andersen1994points,Hind2001pointer}.
An automated \syscall-detection system based on this form of analysis and its impact on the compile-time performance has been demonstrated by Ghavamnia~\etal\cite{Ghavamnia2020}.
This approach also requires changes to the common compilation model, which is not supported by every application.

Hence, as we want to avoid changes to the compilation model and given that our application can tolerate some imprecision, we do not use sophisticated pointer analysis in our prototype implementation and opt for a function-signature based heuristic to determine possible call targets.
Every function in the program where the function type of the call site matches the function type of the definition is considered a possible call target.
Note that, for correctly typed programs, this heuristic is an overapproximation of the actual possible call targets, which corresponds to permitting more \syscalls than are actually needed (\cf \cref{sec:eval-overapproximation}).

Both the LLVM IR passes and the linker are involved in mapping \syscall numbers to functions.
Our analysis pass traverses over all defined functions within the module. It extracts their function signature, functions that are directly called, the function signatures for indirect call sites, and the functions that are referenced by the code (for which function pointers exist, \ie functions that have their address taken).
The latter is similar to what LLVM uses in its implementation of software-based CFI.
This gives rise to the assumption that the resulting call graph is precise enough as applications that use software-based CFI would otherwise not work correctly.

Note that we perform our analysis in the same traversal where we also locate the used \syscall numbers (see \COne), meaning that a single pass over the IR is sufficient.
As function aliases are widely used in musl, we also support them by treating them like copies of the original function.
Finally, references to functions in global initializers are extracted.
In musl, \eg global file structures are used on which functions for reading, writing, and seeking are registered.

Our IR transformation pass serializes the information collected from the analysis pass into a note section of the emitted ELF object for the linker to use this information.
To simplify inspection of the extracted data, we use human-readable JSON as encoding format. %
For a production-ready compiler, binary encoding is preferable to reduce performance overheads. 

\paragrabf{\CThree: Generate \Syscall Set}
By solving challenges \COne and \CTwo, we generate object files containing the serialized \syscall and call graph information.
The linker extracts this information from all the provided input files to perform the actual call graph construction and \syscall number propagation.
Finally, the linker can either generate the set of relevant \syscalls for the application or a flattened call graph for further processing.

In more detail, after loading the call graph metadata, all reachable functions are resolved according to their symbol's linkage specification (\eg local or global, strong or weak), and a list of indirect callable functions is generated.
In the next step, a call graph is constructed in which each node represents a function, and each directed edge represents a possible control-flow transfer from the caller to the callee.
The linker transforms this call graph into a directed acyclic graph (DAG) using Tarjan's algorithm~\cite{tarjan1972depth}, enabling efficient propagation of the information.
Namely, each graph node has to be updated only once by visiting the DAG in post-order.
Using the discovered strongly coupled components, circular call dependencies can be directly resolved by merging the information from all functions that are part of the respective cycle in the original call graph.
As a result, the linker has access to a flattened call graph in which, for every function in the program, all reachable \syscall numbers are known.

Using the flattened call graph, we determine which \syscalls our final application needs.
If we build a static binary, we extract all \syscalls that can be reached from the \textit{main} and the \textit{exit} function and embed them as a note containing a simple list of numbers into the final ELF binary.
For dynamic binaries or shared libraries, we instead embed the flattened call graph, again serialized using JSON as encoding, into a new note of the linked binary for further processing.

\paragrabf{\CFour: Install Seccomp Filters}
After linking with the \CompilerName toolchain, the binary contains annotations containing the application's used \syscall numbers directly or its flattened call graph that still needs to be combined with the additional dynamic libraries.
For static linkage, we delegate the processing to the application itself by additionally linking against our \LibraryName library.
This library contains a constructor that extracts the \syscall numbers and installs the seccomp filters using libseccomp~\cite{Edge2012libseccomp} before the application starts executing.

In the second case, dynamic linkage, we provide two options.
\RewriteTool extracts the embedded call graph from all library dependencies and determines all \syscalls from functions that are reachable from the \textit{main} and \textit{exit} function.
Finally, the tool adds a new note section with information on \syscall numbers.
As the compiler has generated the dynamic binary, we can already link \LibraryName against it automatically.
\SandboxTool performs the same steps except that it does not modify the binary but creates a launcher that sets up the filters before executing the actual binary.

\subsection{Binary Syscall Extraction} \label{sec:binary-extraction}
The second static approach of \ToolName, \BinName, works on the binary level.
While there is less semantic information available than on the compiler level, \BinName works without access to the source code and even for stripped binaries.
In contrast to previous work~\cite{DeMarinis2020}, we demonstrate that the requirement of a PIC binary is not necessary.

\paragrabf{\COne: Identify \Syscall Numbers}
The \syscall number specifying the type of \syscall is not encoded in the \syscall instruction itself.
Instead, the \syscall number is provided by a general-purpose register, \ie \texttt{rax} on x86\_64 or \texttt{x8} on AArch64.
Hence, \BinName has to reconstruct the \syscall number by inferring the content of this register.
This reconstruction results in a list of all \syscalls and their virtual addresses. %

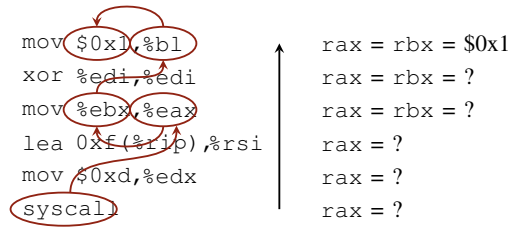
\begin{figure}
  \centering
  \resizebox{0.8\hsize}{!}{
    \input{images/symbolic_backward_execution.tikz}
  }
 \caption{Symbolic backward execution starts from the \syscall instructions and finds the \syscall number by symbolically tracking the corresponding CPU register.}
 \label{fig:backward-execution}
\end{figure}

\BinName uses the capstone framework~\cite{Quynh2014capstone} to disassemble a binary as this framework supports various ISAs, \eg x86\_64 and AArch64.
The disassembly is used as the base for identifying \syscall numbers.
Starting from a \syscall instruction, \BinName leverages symbolic backward execution~\cite{Ma2011directed}.
Tracking back from the \syscall instruction, \BinName tracks the register's symbolic value containing the \syscall number.
In many cases, the immediate for the \syscall number is directly moved to the register before the \syscall instruction as it is a constant value.
However, in some cases, there is at least some form of register-to-register transfer involved.
These transfers also include register copies where only a lower part of the register is involved.
Thus, as illustrated in \Cref{fig:backward-execution}, \BinName keeps the content of the register symbolic and steps back through the binary, symbolically evaluating operations.
This symbolic backward execution is repeated until either a concrete immediate for the \syscall number is identified, or after a user-definable number of instructions have been analyzed without successfully identifying the immediate.\footnote{For the evaluation, we set this number to 30, which was sufficiently large.}
One failure reason can be that the \syscall instruction is a call or jump target, \ie there are potentially multiple call sites reaching the instruction with different \syscall numbers.
Such a situation would require a more complex symbolic execution. %
Luckily, the \syscall instruction is usually inlined, and thus, we do not consider such situations for our proof of concept.

\begin{figure}
\centering
\resizebox{0.8\hsize}{!}{
 \input{images/syscall_mapping.tikz}
 }
 \caption{\BinName creates an $n:m$ mapping between exported functions (ellipse) and \syscalls (rectangle).} %
 \label{fig:cfg-bin-analysis}
\end{figure}
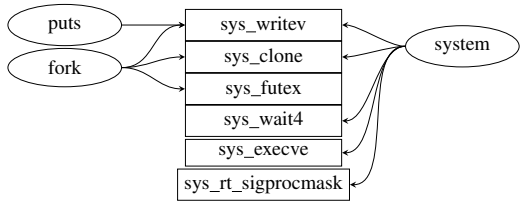

\paragrabf{\CTwo: Reconstruct \Syscall Call Sites}
To reduce the overapproximation of used \syscalls, \BinName analyzes the binary's control-flow graph (CFG) to map identified \syscalls to (exported) functions.
After this analysis, \BinName has a set of all possible \syscalls per exported function (\cf~\Cref{fig:cfg-bin-analysis}).

We rely on angr~\cite{Wang2017angr} to statically create a CFG of the binary.
Based on the basic blocks of all functions in the CFG, we assign every \syscall identified in \COne to a function.
Such a function might not be an exported function or even a named function, but any block identified as a function.

\BinName traverses the CFG from each exported function as the root node to identify reachable functions with a \syscall instruction.
Assuming a correctly reconstructed CFG from angr and correctly identified \syscall numbers (\COne), this yields a set of possible \syscalls per exported function (\cf \Cref{fig:cfg-bin-analysis}).

\paragrabf{\CThree: Generate \Syscall Set}
To solve challenge \CThree, we have to combine the information created from solving \CTwo for all binaries, \ie the application binary and all of its dynamically-linked libraries.
We cannot create a complete CFG over the application binary and its libraries as this would take multiple hours to days, depending on the size of the application and the number of dynamic libraries.
As a tradeoff, we chose to overapproximate the number of possible \syscalls by relying on individual CFGs that we merge.
We only consider functions that are defined in the dynamic symbol table of the ELF file.
These functions are found in the dynamic libraries loaded by the application.
Hence, we search for these functions in the shared object dependencies and look up the used \syscalls for the function in the respective library.
Note that shared libraries can also have a dynamic symbol table if they call functions from other libraries.
Thus, this process is repeated for all dynamic symbols of all shared object dependencies.

Solving challenge \CThree yields a set of \syscalls that the application can potentially call.
This assumes that no dynamic libraries are loaded at runtime, \eg via \texttt{dlopen}, and that the application does not execute a different binary at runtime, \eg via \texttt{exec}.
In such cases, we would have to resort to \PTwo, as the complete set of \syscalls cannot be determined statically.

\paragrabf{\CFour: Install Seccomp Filters}
From the complete set of \syscalls, \BinName has to create filter rules and apply them to the binary.
We cannot simply compile the filters with the application (\cf \CompilerName).
Instead, \BinName supports two different variants: binary rewriting, and building a sandbox wrapper (\cf \Cref{fig:overview}).
\SandboxTool is a simple application that sets up the filter rules and starts the target application.

With binary rewriting, \BinName stores the \syscall numbers in the ELF binary and injects a new shared object dependency, \LibraryName.
The library provides a constructor function, which is called before the actual application starts.
In the constructor function, the library parses the filters stored in the binary to apply the seccomp filter rules.
The advantage of a rewritten binary is that it does not need any launcher application.

\section{Dynamic Refinement} \label{sec:dynamic-sandboxing}
In this section, we discuss the optional \PTwo component \DynamicName, a method to dynamically refine the previously detected \syscall filters.
The dynamic approach simplifies the challenges \COne to \CFour by inspecting \syscalls just-in-time in a secure and controlled environment during development.

\subsection{Limitations of Static Approaches}
While our approach for statically detecting an application's \syscalls works well for most binaries (\cf \Cref{sec:evaluation}), there are inherent limitations to a static approach.
Dynamically loaded libraries, \eg codecs, plugins, self-modifying, or just-in-time-compiled code, often cannot be analyzed statically.

Moreover, the current implementation of seccomp is not flexible enough to handle scenarios involving child processes with a different set of \syscalls, as a child inherits its parent's filters and can only further restrict but not relax them.
For a child to work as intended, the parent also needs to install a seccomp filter for the \syscalls the child uses as otherwise the operating system kills the child.

\subsection{Implementation Details}

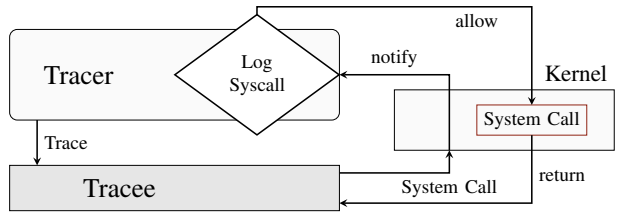
\begin{figure}[t]
  \centering
  \resizebox{\hsize}{!}{
    \input{images/dynamic-trace.tikz}
  }
  \caption{The tracer gets notified by the kernel when the tracee executes a \syscall.
  The tracer logs the \syscall and informs the kernel to execute the \syscall.
  }
  \label{fig:dynamic-workflow-sandboxing}
\end{figure}

In our prototype, \DynamicName is a strace-like \syscall-tracing component linked against the target application or used as a standalone wrapper for a binary (\cf \Cref{fig:overview}).
This allows \DynamicName to work with \BinName and \CompilerName. %
If desired, it can also be used without any of the two static components to identify the required \syscalls.

In either case, \DynamicName, \ie the \textit{tracer}, first creates a child process, the \textit{tracee}.
\DynamicName then installs seccomp filters for all \syscalls in a way that informs the tracer about a seccomp violation.
To enable this behavior, the tracer needs to attach itself to the tracee. %
The tracee then stops execution until it receives the continue signal from the tracer to ensure that it successfully attached itself.
If the child process creates a new child process, the tracer is automatically attached to the newly created child process.
The tracer is then also informed of the unsuccessful execution of the child's \syscalls.

Upon receiving the notification of a violating \syscall, \DynamicName extracts the \syscall number (\COne), logs it (\CThree), and allows it for all future occurrences (\cf \cref{sec:high-level}).
As the \syscall is indeed executed, it is inherent that it is reachable (\CTwo).
We illustrate this whole process in \Cref{fig:dynamic-workflow-sandboxing}.

Once \DynamicName has finished tracing the application, it cross-references the list of obtained \syscalls with the ones obtained in \POne.
If a \syscall is missing, it modifies the allowlist to include the newly detected \syscall.
Optionally, it can also be used to remove \syscalls that \POne identified but which were never executed during \PTwo.

\section{Evaluation}\label{sec:evaluation}
In this section, we evaluate the performance, functional correctness, and security of \ToolName.
Our evaluation is in line with related work~\cite{Ghavamnia2020,DeMarinis2020} while improving on it in several points, \ie we evaluate several parts that were omitted by these works.
In the performance evaluation, we evaluate the one-time overheads of \ToolName, such as compile time and binary-analysis time. %
We also discuss the runtime overhead seccomp introduces. %
For the functional correctness, we evaluate whether \ToolName causes any issues in terms of functionality of existing real-world software, \eg crashes. %
We also perform a 6 months long evaluation of an Nginx server with its \syscall interface restricted by \CompilerName.
In the security evaluation, we evaluate the ability of \ToolName to prevent the dangerous \texttt{exec} \syscall and the overapproximation of \syscalls in general.
With the latter, we are the first to demonstrate how tight the automatically generated filters are.
Furthermore, we evaluate how well \ToolName can mitigate real-world exploits. %
Finally, we discuss related works in this field to show the main differences to our work.

\subsection{Setup}
For the evaluation of \ToolName, we focus on x86\_64.
Note that the only architecture-dependent part of \ToolName is the extraction of the \syscall number.
Hence, we do not expect significant differences for other architectures.
We also verified that the general approach works across architectures by successfully extracting the \syscall numbers from musl libc for both x86\_64 and AArch64.

We evaluate \ToolName on various real-world applications shown in \cref{tbl:syscalls-evaluation}, including client, server, and database software.
While busybox may be seen as a non-obvious choice, it is in line with previous work that used coreutils for the evaluation~\cite{quach2018debloating}.
We instead chose busybox as the number of provided utilities is 3 times higher, making it a better choice for our evaluations.
For evaluating \CompilerName, we compile the binaries statically with and without \ToolName enabled using our modified compiler.
For \BinName, we compile the applications dynamically using GCC 7.5.0-3 on Ubuntu 18.04.4. %
For the sake of brevity, we do not evaluate every combination of components and sandboxes but focus on \LibraryName for \CompilerName and \SandboxTool for \BinName.

\begin{table*}[t]
  \centering
  \resizebox{\hsize}{!}{
    \begin{tabular}{llccrrrrcccccccc}
       & \textbf{Software} & \multicolumn{2}{c}{\textbf{\makecell{\#Syscalls                                                                                                                                                                                                                                                                                                                                     \\Found / Used / \PTwo Added}}} & \multicolumn{2}{c}{\textbf{Size                                                                                                                                                                                                         Overhead}} & \multicolumn{2}{c}{\textbf{Analysis Time}} & \multicolumn{2}{c}{\textbf{\texttt{exec}}} & \multicolumn{2}{c}{\textbf{\texttt{mprotect}}} & \multicolumn{2}{c}{\textbf{\makecell{Fully\\ Mitigated}}} & \multicolumn{2}{c}{\textbf{\makecell{Subvariant\\ Mitigated}}} \\
       &                   & \srcfile                                           & \binfile                        & Compiler (\srcfile)                         & Binary (\binfile)                       & Compiler (\srcfile)                        & Binary (\binfile)   & \srcfile & \binfile & \srcfile & \binfile & \srcfile            & \binfile            & \srcfile            & \binfile            \\

      \toprule
      \multirow{13}{*}{\,\raisebox{10pt}{\tikz{\draw [decorate,decoration={brace,amplitude=5pt}] (0,0.5) -- node[midway,above,rotate=90,yshift=0.15cm]{Client} (0,5.0);}}}
       & ls                & \SIx{24} / \SIx{14} / \SIx{0}                      & \SIx{39} / \SIx{18} / \SIx{1}   & +\SI{173}{\kilo\byte} (\SI{253}{\percent})  & +\SI{288}{\byte} (\SI{1.08}{\percent})  & +\SI{0.38}{\second} (\SI{1.72}{\percent})  & \SI{3.041}{\second} & \cmark   & \cmark   & \cmark   & \xmark   & \SI{81.1}{\percent} & \SI{81.1}{\percent} & \SI{87.5}{\percent} & \SI{87.5}{\percent} \\
       & chown             & \SIx{22} / \SIx{11} / \SIx{0}                      & \SIx{36} / \SIx{14} / \SIx{0}   & +\SI{174}{\kilo\byte} (\SI{369}{\percent})  & +\SI{280}{\byte} (\SI{1.52}{\percent})  & +\SI{0.29}{\second} (\SI{1.35}{\percent})  & \SI{2.777}{\second} & \cmark   & \cmark   & \cmark   & \xmark   & \SI{81.7}{\percent} & \SI{81.7}{\percent} & \SI{87.8}{\percent} & \SI{87.8}{\percent} \\
       & cat               & \SIx{18} /  \phantom{0}\SIx{6} / \SIx{0}           & \SIx{34} / \SIx{13} / \SIx{0}   & +\SI{174}{\kilo\byte} (\SI{397}{\percent})  & +\SI{272}{\byte} (\SI{1.9}{\percent})   & +\SI{0.08}{\second} (\SI{2.29}{\percent})  & \SI{2.576}{\second} & \cmark   & \cmark   & \cmark   & \xmark   & \SI{82.3}{\percent} & \SI{81.7}{\percent} & \SI{88.1}{\percent} & \SI{87.8}{\percent} \\
       & pwd               & \SIx{16} /  \phantom{0}\SIx{4} / \SIx{0}           & \SIx{34} / \SIx{14} / \SIx{0}   & +\SI{175}{\kilo\byte} (\SI{430}{\percent})  & +\SI{272}{\byte} (\SI{1.92}{\percent})  & +\SI{0.21}{\second} (\SI{0.98}{\percent})  & \SI{2.507}{\second} & \cmark   & \cmark   & \cmark   & \xmark   & \SI{85.1}{\percent} & \SI{81.7}{\percent} & \SI{90.3}{\percent} & \SI{87.8}{\percent} \\
       & diff              & \SIx{25} / \phantom{0}\SIx{9} / \SIx{0}            & \SIx{36} /  \SIx{16} / \SIx{0}  & +\SI{173}{\kilo\byte} (\SI{304}{\percent})  & +\SI{280}{\byte} (\SI{1.25}{\percent})  & +\SI{0.06}{\second} (\SI{1.44}{\percent})  & \SI{2.946}{\second} & \cmark   & \cmark   & \cmark   & \xmark   & \SI{81.1}{\percent} & \SI{81.7}{\percent} & \SI{87.5}{\percent} & \SI{87.8}{\percent} \\
       & dmesg             & \SIx{15} /  \phantom{0}\SIx{5} / \SIx{0}           & \SIx{34} / \SIx{14} / \SIx{0}   & +\SI{176}{\kilo\byte} (\SI{439}{\percent})  & +\SI{272}{\byte} (\SI{1.92}{\percent})  & +\SI{0.08}{\second} (\SI{2.17}{\percent})  & \SI{2.452}{\second} & \cmark   & \cmark   & \cmark   & \xmark   & \SI{85.1}{\percent} & \SI{81.7}{\percent} & \SI{90.3}{\percent} & \SI{87.8}{\percent} \\
       & env               & \SIx{15} / \phantom{0}\SIx{3} / \SIx{0}            & \SIx{33} /  \SIx{13} / \SIx{0}  & +\SI{175}{\kilo\byte} (\SI{416}{\percent})  & +\SI{272}{\byte} (\SI{1.92}{\percent})  & +\SI{0.07}{\second} (\SI{1.88}{\percent})  & \SI{2.416}{\second} & \xmark   & \cmark   & \cmark   & \xmark   & \SI{81.7}{\percent} & \SI{81.7}{\percent} & \SI{88.4}{\percent} & \SI{87.8}{\percent} \\
       & grep              & \SIx{20} / \SIx{11} / \SIx{0}                      & \SIx{34} / \SIx{16} / \SIx{0}   & +\SI{174}{\kilo\byte} (\SI{177}{\percent})  & +\SI{272}{\byte} (\SI{1.49}{\percent})  & +\SI{0.41}{\second} (\SI{1.88}{\percent})  & \SI{2.748}{\second} & \cmark   & \cmark   & \cmark   & \xmark   & \SI{81.7}{\percent} & \SI{81.7}{\percent} & \SI{87.8}{\percent} & \SI{87.8}{\percent} \\
       & true              & \phantom{0}\SIx{3} /  \phantom{0}\SIx{1} / \SIx{0} & \SIx{32} / \SIx{12} / \SIx{0}   & +\SI{200}{\kilo\byte} (\SI{3277}{\percent}) & +\SI{264}{\byte} (\SI{4.4}{\percent})   & +\SI{0.09}{\second} (\SI{2.55}{\percent})  & \SI{9.908}{\second} & \cmark   & \cmark   & \cmark   & \xmark   & \SI{98.3}{\percent} & \SI{83.4}{\percent} & \SI{98.4}{\percent} & \SI{88.7}{\percent} \\
       & head              & \SIx{17} / \phantom{0}\SIx{7} / \SIx{0}            & \SIx{33} / \SIx{13} / \SIx{0}   & +\SI{174}{\kilo\byte} (\SI{434}{\percent})  & +\SI{272}{\byte} (\SI{1.92}{\percent})  & +\SI{0.06}{\second} (\SI{1.64}{\percent})  & \SI{2.436}{\second} & \cmark   & \cmark   & \cmark   & \xmark   & \SI{82.3}{\percent} & \SI{81.7}{\percent} & \SI{88.1}{\percent} & \SI{87.8}{\percent} \\
      \cmidrule{2-16}
       & git               & \SIx{82} / \SIx{42} / \SIx{1}                      & \SIx{85} / \SIx{42} / \SIx{2}   & +\SI{219}{\kilo\byte} (\SI{4.5}{\percent})  & +\SI{448}{\byte} (\SI{0.003}{\percent}) & +\SI{18.5}{\second} (\SI{28.18}{\percent}) & \SI{247}{\second}   & \xmark   & \xmark   & \xmark   & \xmark   & \SI{34.3}{\percent} & \SI{58.3}{\percent} & \SI{55.9}{\percent} & \SI{73.4}{\percent} \\
      \cmidrule{2-16}
       & FFmpeg            & \SIx{63} / \SIx{27} / \SIx{1}                      & \SIx{91} / \SIx{27} / \SIx{2}   & +\SI{190}{\kilo\byte} (\SI{0.21}{\percent}) & +\SI{472}{\byte} (\SI{0}{\percent})     & +\SI{268}{\second} (\SI{27.14}{\percent})  & \SI{643}{\second}   & \xmark   & \cmark   & \xmark   & \xmark   & \SI{33.1}{\percent} & \SI{34.9}{\percent} & \SI{57.8}{\percent} & \SI{44.1}{\percent} \\
      \cmidrule{2-16}
       & mutool            & \SIx{61} / \SIx{16} / \SIx{1}                      & \SIx{69} / \SIx{15} / \SIx{0}   & +\SI{189}{\kilo\byte} (\SI{0.48}{\percent}) & +\SI{376}{\byte} (\SI{0.001}{\percent}) & +\SI{3.17}{\second} (\SI{0.69}{\percent})  & \SI{164}{\second}   & \xmark   & \cmark   & \xmark   & \xmark   & \SI{52.0}{\percent} & \SI{38.3}{\percent} & \SI{69.4}{\percent} & \SI{60.3}{\percent} \\
      \cmidrule{2-16}
       & memcached         & \SIx{88} / \SIx{54} / \SIx{1}                      & \SIx{102} / \SIx{59} / \SIx{4}  & +\SI{216}{\kilo\byte} (\SI{28.9}{\percent}) & +\SI{456}{\byte} (\SI{0.13}{\percent})  & +\SI{0.35}{\second} (\SI{5.5}{\percent})   & \SI{8}{\second}     & \xmark   & \cmark   & \xmark   & \xmark   & \SI{30.3}{\percent} & \SI{33.7}{\percent} & \SI{50.0}{\percent} & \SI{41.9}{\percent} \\
      \midrule
      \multirow{2}{*}{\,\raisebox{10pt}{\tikz{\draw [decorate,decoration={brace,amplitude=5pt}] (0,0.1) -- node[midway,above,rotate=90,yshift=0.15cm]{DB} (0,0.75);}}}
       & redis-server      & \SIx{85} / \SIx{35} / \SIx{1}                      & \SIx{93} / \SIx{42} / \SIx{3}   & +\SI{216}{\kilo\byte} (\SI{11.2}{\percent}) & +\SI{472}{\byte} (\SI{0}{\percent})     & +\SI{2.4}{\second} (\SI{2.7}{\percent})    & \SI{41}{\second}    & \xmark   & \xmark   & \xmark   & \xmark   & \SI{30.3}{\percent} & \SI{32.0}{\percent} & \SI{54.1}{\percent} & \SI{54.4}{\percent} \\
      \cmidrule{2-16}
       & sqlite3           & \SIx{92} / \SIx{72} / \SIx{1}                      & \SIx{102} / \SIx{72} / \SIx{13} & +\SI{215}{\kilo\byte} (\SI{5.9}{\percent})  & +\SI{456}{\byte} (\SI{0.02}{\percent})  & +\SI{0.8}{\second} (\SI{7.2}{\percent})    & \SI{45}{\second}    & \xmark   & \xmark   & \xmark   & \xmark   & \SI{62.9}{\percent} & \SI{32.6}{\percent} & \SI{75.3}{\percent} & \SI{57.5}{\percent} \\
      \midrule
      \multirow{2}{*}{\,\raisebox{10pt}{\tikz{\draw [decorate,decoration={brace,amplitude=5pt}] (0,0.1) -- node[midway,above,rotate=90,yshift=0.15cm]{Server} (0,0.75);}}}
       & Nginx             & \SIx{105} / \SIx{48} / \SIx{0}                     & \SIx{106} / \SIx{51} / \SIx{4}  & +\SI{217}{\kilo\byte} (\SI{1.5}{\percent})  & +\SI{528}{\byte} (\SI{0.003}{\percent}) & +\SI{7.9}{\second} (\SI{10.53}{\percent})  & \SI{277}{\second}   & \xmark   & \cmark   & \cmark   & \xmark   & \SI{32.0}{\percent} & \SI{30.9}{\percent} & \SI{38.8}{\percent} & \SI{40.0}{\percent} \\
      \cmidrule{2-16}
       & httpd             & \SIx{98} / \SIx{50} / \SIx{1}                      & \SIx{106} / \SIx{46} / \SIx{0}  & +\SI{218}{\kilo\byte} (\SI{8.3}{\percent})  & +\SI{504}{\byte} (\SI{0.04}{\percent})  & +\SI{4.1}{\second} (\SI{5}{\percent})      & \SI{16.8}{\second}  & \xmark   & \xmark   & \xmark   & \xmark   & \SI{29.7}{\percent} & \SI{30.3}{\percent} & \SI{50.9}{\percent} & \SI{44.4}{\percent} \\
      \bottomrule
    \end{tabular}
  }
  \caption{Results for the compiler- (\protect\srcfile) and binary-based (\protect\binfile) approach of \ToolName, respectively.
  For each software, we show the number of detected \syscalls in \POne, used \syscalls, and added \syscalls in \PTwo, the size overhead of the annotations, compile-time overhead (for \CompilerName), and binary analysis time (for \BinName).
  The \texttt{exec} and \texttt{mprotect} columns indicate whether \ToolName blocks (\cmark) the \texttt{execve} and \texttt{execveat} or \texttt{mprotect} \syscalls respectively.
  We also show the percentage of fully mitigated CVEs and individual subvariants of these CVEs targeting the kernel.
  Note that the \syscalls added in \PTwo are only necessary in our proof-of-concept implementation of \ToolName as they appear in edge cases, which can be handled in a production-ready implementation.}
  \label{tbl:syscalls-evaluation}
\end{table*}

\begin{table}[t]
  \centering
  \adjustbox{max width=\hsize}{
    \begin{tabular}{lrrr}
      \textbf{Shared library} & \textbf{Vanilla}     & \textbf{Annotated}    & \textbf{Overhead}    \\
      \toprule
      musl libc.so            & \SI{815}{\kilo\byte} & \SI{1007}{\kilo\byte} & \SI{23.63}{\percent} \\
      libssl.so               & \SI{657}{\kilo\byte} & \SI{1.7}{\mega\byte}  & \SI{161}{\percent}   \\
      libcrypto.so            & \SI{4.1}{\mega\byte} & \SI{23}{\mega\byte}   & \SI{460}{\percent}   \\
      \bottomrule
    \end{tabular}
    }
  \caption{We evaluate the size overhead of the compiler-based approach of \ToolName on shared libraries.
    The overhead is based on a vanilla version of the respective shared library.}
  \label{tbl:shared-library-size}
\end{table}

\subsection{Performance Evaluation}
In this section, we evaluate the performance of \ToolName.
This includes the one-time overheads for compiling (\Cref{sec:eval:perf:compile}) or binary analysis (\Cref{sec:eval:perf:binary}), the increase in binary size (\Cref{sec:eval:perf:size}), and runtime overheads (\Cref{sec:eval:perf:runtime}).

\subsubsection{Compile-Time Overhead}\label{sec:eval:perf:compile}
We analyze the impact \CompilerName has on the compile time of an application.
To make comparison possible, we compile the application \SIx{10} times with and without our modification enabled, always using our modified compiler, and use the average compile time over these runs.

As the results show, we observe the worst-case overhead for the git application with an increase from an average of \SI{65.5}{\second} ($\sigma_{\bar x}=\SIx{0.094}$, $N=10$) to \SI{84}{\second} ($\sigma_{\bar x}=\SIx{0.054}$, $N=10$), an increase of \SI{28}{\percent}.
For the busybox utilities combined, the average increases from \SI{10.94}{\second} ($\sigma_{\bar x}=\SIx{2.88}$, $N=\SIx{10}$) to \SI{10.99}{\second} ($\sigma_{\bar x}=\SIx{2.79}$, $N=\SIx{10}$).
When compared to related work~\cite{Ghavamnia2020}, we observe a speedup of factor \SIx{73} for Nginx when using \CompilerName.
This low overhead makes it a feasible approach to be used in everyday development cycles.

\subsubsection{Binary Extraction Runtime}\label{sec:eval:perf:binary}
For \BinName, we evaluate the time it takes for extracting the \syscalls from the dynamic binary.
We assume that default dependencies like \textit{libc.so} have already been processed and that their extracted call graph is available to the user.
For completeness, we timed the extraction of \syscalls from \textit{libc.so}, which takes on average \SI{44.66}{\second} ($\sigma_{\bar x}=\SIx{0.18}$, $N=\SIx{10}$).
For the applications themselves, we can see in \cref{tbl:syscalls-evaluation} that the extraction process is in the range of \SIx{2} to \SI{10}{\second} for the individual busybox utilities, with an average time of \SI{3.4}{\second} ($\sigma_{\bar x}=\SIx{0.73}$, $N=\SIx{10}$).
For large binaries like FFmpeg ($>$ \SI{100}{\mega\byte}) and its dynamic dependencies, the extraction takes around \SI{11}{\minute}.

\subsubsection{Binary Size Analysis}\label{sec:eval:perf:size}
The code size does not increase with adding filters, only the binary size increases by the meta-information.
Moreover, \LibraryName and libseccomp are potentially linked to the application.

\paragrabf{Compiler}
We analyze the size of the binary produced by \CompilerName compared to a vanilla application.
\ToolName needs to treat static and dynamic ELF files differently as \syscall numbers of externally linked libraries are not known.
In a static binary, we only add the set of \syscall numbers to the binary and link against \LibraryName and libseccomp.
As both libraries are of fixed size, the maximum overhead in a static binary is limited by the number of \syscalls Linux provides, \ie \SIx{349} on Linux 5.0.
\cref{tbl:syscalls-evaluation} shows the overhead for statically linked binaries.
As expected, the overhead is quite small in large binaries, \eg FFmpeg. %
In the small busybox utilities, the overhead appears to be huge ($>$ \SI{177}{\percent}), but as these binaries sizes are in the lower kilobyte range (\SIx{40}-\SI{100}{\kilo\byte}), linking against two additional libraries drastically increases the size.
Nevertheless, the binaries remain in the kilobyte range.

For dynamic binaries and shared libraries, we have to embed the entire call graph as we need the information later on to determine the required \syscalls.
\cref{tbl:shared-library-size} shows the size increase for three shared libraries.
In \textit{libcrypto.so}, we observe a worst-case increase from \SI{4.1}{\mega\byte} to \SI{23}{\mega\byte} (\SI{460}{\percent}).
The overhead also increases with the size of the binary as the call graph is larger for the larger codebase.

\paragrabf{Binary}
\cref{tbl:syscalls-evaluation} shows the increase for \BinName.
We opted to generate a binary that needs to be launched by \SandboxTool instead of rewriting the binary. %
Still, for simplicity, we embed the detected \syscalls in the binary from where our wrapper extracts the information.
As we embed only the numbers, the overhead in all \SIx{18} applications is less than  \SI{2}{\percent}.
Binary rewriting incurs the additional overhead of adding the dependency on \LibraryName and libseccomp, but this increase is again insignificant and similar to the observed result of \CompilerName.

\subsubsection{Runtime Overhead and Seccomp}\label{sec:eval:perf:runtime}
For the static approaches, the only overhead compared to manually crafted seccomp filters is the parsing of the \syscall numbers. %
As this is done during application startup, it is a one-time overhead that depends on the number of rules that need to be set up.
Hence, we investigate the overhead for setting up the application with the smallest (\textit{true}) and largest (\textit{Nginx}) number of \syscalls based on \CompilerName.
For Nginx, the setup time takes on average \SI{9.92}{\milli\second} ($\sigma_{\bar x}=\SIx{0.007}$, $N=\SIx{10000}$) while it only takes \SI{0.58}{\milli\second} ($\sigma_{\bar x}=\SIx{0.004}$, $N=\SIx{10000}$) for \textit{true}.
The remaining slowdown is then introduced by seccomp itself, which is unavoidable if a developer decides to sandbox an application with it.
Previous work has shown that its performance depends on the number of filters that are installed and the rule's position within the filter list~\cite{Hromatka2018,Tizen2014}.
The Linux developers are currently working on improving the performance of seccomp~\cite{Corbet2020syscall}.

\subsubsection{Dynamic Refinement Overhead}
\label{sec:dynamic-overhead}
As a microbenchmark, we analyze the impact of \DynamicName on the \syscall latency.
We first benchmark the latency of the \textit{getppid} \syscall without \DynamicName in place \SIx{1} million times.
The latency of \textit{getppid} on our test system (Ubuntu 18.04.4, kernel 5.0.21-050021-generic) is \SIx{1358} ($\sigma_{\bar x}=\SIx{0.91}$, $N=\SIx{1000000}$) cycles.
With \DynamicName, we observe an average latency of \SI{17103} ($\sigma_{\bar x}=\SIx{5.52}$, $N=\SIx{1000000}$) cycles, an increase of approximately \SI{1160}{\percent}.
While this increase seems large, it is supposed to be used as an optional step during development.
Hence, we consider this to be less of a problem as it does not impact the released application.

\subsection{Functional-Correctness Evaluation}\label{sec:correctness}
A critical aspect of \ToolName is that the sandboxed binaries still work as intended without observing crashes.
Related work~\cite{Ghavamnia2020} tested each application \SIx{100} times using various workloads. %
For a fair comparison, we perform the same tests. 
Moreover, for those applications where a test suite is available, we execute these test suites to reach higher coverage, ensuring that we do not miss edge cases. 
Beyond previous work~\cite{Ghavamnia2020,DeMarinis2020}, we also extend our evaluation with code coverage results to show that large parts of the application are actually executed.
We also perform a 6 month long test of Nginx sandboxed by \CompilerName.

\begin{table}[t]
  \centering
  \adjustbox{max width=\hsize}{
    \footnotesize
    \begin{tabular}{lrrr}
      \textbf{Software} & \textbf{\makecell{Coverage\\ Lines}} & \textbf{\makecell{Coverage\\ Functions}}\\
      \toprule
      FFmpeg            & \SI{59.3}{\percent}        & \SI{61.7}{\percent} \\
      memcached         & \SI{77}{\percent}          & \SI{91.9}{\percent} \\
      redis             & \SI{77}{\percent}          & \SI{61.5}{\percent} \\
      \bottomrule
    \end{tabular}
  }
  \caption{Coverage results for selected applications.}
  \label{tbl:coverage}
\end{table}

In more detail, we first apply \ToolName to the binaries shown in \cref{tbl:syscalls-evaluation}. %
Note that obtaining a sound ground truth of whether all \syscalls are detected is infeasible and would require time-consuming formal proofs that are out-of-scope for this paper.
Hence, we rely on executing the available test suites that should cover many of the different code paths available in the tested application.
This is, for instance, possible for FFmpeg, memcached, redis, Nginx, and sqlite3.
In other cases, we execute the binaries with different configurations to ensure that we reach as many different code paths as possible, similar to what related work has done~\cite{Ghavamnia2020}.
We observed no crashes in applications sandboxed with \ToolName. %
Even if a \syscall is missed in \POne, \PTwo can be used to add it, ensuring correct functionality.

While this is not an exhaustive test, it can be assumed that test suites for large applications are designed for complete functionality coverage and thorough testing of critical components in particular.
Based on the latter, it is a reasonable assumption that our functional-correctness test tests whether all \syscalls in the core functionality of the tested application are found.
To further substantiate this, we perform a coverage test for a selection of applications shown in \cref{tbl:syscalls-evaluation}.
We show the result of these coverage tests in \cref{tbl:coverage}.
Additionally to our results, the sqlite developers always maintain \SI{100}{\percent} branch and \SI{100}{\percent} MC/DC coverage~\cite{sqlite2020testing}.
While not perfect, the results indicate that large parts of the respective applications are executed and, to a certain degree, demonstrate the functional correctness of the applications after \ToolName has been applied.
In future work, we would like to employ coverage-guided fuzzing to better estimate whether all required \syscalls are found.

Programs using \texttt{fork}+\texttt{exec}, \eg \textit{git-diff}, exhibit the inherent problem of seccomp, namely that a child program inherits its parent's filters.
If the child uses a \syscall blocked by the parent, the child crashes.
For such applications, \PTwo is necessary to ensure functionality.
Out of the \SIx{18} tested applications, \PTwo was only necessary for two of them, namely \textit{git-diff} and \textit{git-log} as they performed \syscalls blocked by their parent.
After refining the filters using \DynamicName, both successfully completed their task.

\paragraph{Adding Missed \Syscalls using \PTwo.}
We evaluated how many \syscalls the static approaches missed. %
For \CompilerName, \DynamicName added \SIx{4} \syscalls to musl libc, which are then propagated to the individual applications if the corresponding function is used, \eg \textit{clone}.
\cref{tbl:syscalls-evaluation} shows for each application how many \syscalls were added in \PTwo. 

For \BinName and busybox, \PTwo only had to add a \syscall in \textit{ls}.
The largest amount of added \syscalls appears in sqlite3, where \DynamicName adds \SIx{13} \syscalls.

These missed \syscalls are currently only a limitation of our proof-of-concept implementation.
All these \syscalls occur in edge cases that our implementations do not, but a full implementation can cover.
Hence, they do not necessitate \DynamicName to be a mandatory phase of \ToolName.

\paragrabf{Long-Term Study using Nginx}
To further demonstrate the functional correctness of \ToolName, we performed a long-term study of 6 months using Nginx.
In this test, we compiled a static version of Nginx using \CompilerName, which we then deployed to a real-world server to host a website.
The number of blocked \syscalls for that Nginx binary is shown in \cref{tbl:syscalls-evaluation}, \ie \SIx{105}.
Over this period of \SIx{6} months, the server handled around \SIx{100000} requests without ever triggering a seccomp violation.
This further demonstrates that \CompilerName can infer all \syscalls necessary for a successful operation of Nginx on a real-world system.

\subsection{Security Evaluation}
To evaluate how \ToolName increases the security of sandboxed applications, we analyze how often dangerous \syscalls, \eg \texttt{exec}, are blocked (\Cref{sec:eval:sec:danger}), the number of \syscalls not blocked even though they are not used by the application (\Cref{sec:eval-overapproximation}), the number of mitigated real-world exploits (\Cref{sec:eval-security}), and how malicious SGX enclaves can be prevented (\Cref{sec:eval:sec:sgx}).

\subsubsection{Preventing Dangerous \Syscalls}\label{sec:eval:sec:danger}
Three of the more dangerous \syscalls that Linux provides are the two \syscalls in the \texttt{exec} group, \ie \texttt{execve} and \texttt{execveat}, and the \texttt{mprotect} \syscall.
With the \texttt{exec} \syscalls available, an attacker can execute an arbitrary binary in the presence of an exploitable memory safety violation~\cite{Carlini2014COP}.
In fact, most libc versions even contain a ROP gadget that leverages the \texttt{exec} \syscall to open a shell~\cite{Jurczyk2015gadget}.
Hence, an attacker can execute a new program in the context of the current one.
With \texttt{mprotect}, an attacker can modify the permissions of existing memory, \ie make it executable.
While \texttt{mmap} can be used to map memory as executable, we did not consider it in our evaluation.
We consider attacks not relying on \syscalls~\cite{Chen2005data} as out of scope.

Even with \ToolName, certain attacks are still possible, \eg adding an ssh key if a privileged application is exploited and the \textit{open}/\textit{write} \syscalls are allowed.
Note that these attacks are also possible with \ToolName, but other attacks are prevented, improving the overall system security.
Hence, \ToolName improves the status quo, which is the goal of this work.

\paragrabf{Compiler}
We evaluate the effectiveness of \CompilerName in preventing the \texttt{exec} and \texttt{mprotect} \syscalls (\cref{tbl:syscalls-evaluation}).
In busybox, we prevent the \texttt{exec} \syscalls in \SIx{9} out of \SIx{10} cases and \texttt{mprotect} in all \SIx{10}.
Additionally, we also evaluated all the remaining busybox utilities and prevent \texttt{exec} in \SIx{313} out of \SIx{396} (\SI{79.0}{\percent}) of them and \texttt{mprotect} in all \SIx{396} (\SI{100}{\percent}).
In Nginx, we cannot prevent \texttt{exec}, but we do prevent \texttt{mprotect}.
In the other applications, we can prevent neither of them as our compiler detects a potential call to a function that contains the respective \syscalls.

\paragrabf{Binary}
As \cref{tbl:syscalls-evaluation} shows, \BinName prevents the \texttt{exec} \syscalls in all of the shown busybox utilities.
The analysis of the remaining busybox utilities showed that we can also prevent it in all of them.
This result differs from \CompilerName which could not block the \texttt{exec} \syscall in the \textit{env} utility.
We manually verified that the \syscalls are indeed not required in the application.
The \texttt{mprotect} analysis showed the opposite behavior as it is not blocked in any of the applications. %
For Nginx, memcached, mutool, and FFmpeg, we were also able to block the \texttt{exec} \syscalls without crashing the application, but not \texttt{mprotect}.
We could not block either one of them in git, httpd, redis, and sqlite3.
For git and the \texttt{exec} \syscalls, the reason is that some of the git commands rely on other applications to run, \ie the configured pager for commands like \textit{diff} or \textit{log}.
The explanation of why we cannot block \texttt{mprotect} using \BinName is the point of time at which we start blocking \syscalls.
In \CompilerName, we block \syscalls that are reachable only from the \textit{main} and \textit{exit} functions, while we block them from the start of the application in \BinName.
Hence, we need to allow \texttt{mprotect} as it is required for setting up the application.
In a full implementation, the functions necessary for program startup can be removed from the analysis, potentially removing the \textit{mprotect} \syscall.

\subsubsection{Overapproximation of \Syscalls}\label{sec:eval-overapproximation}
As \cref{tbl:syscalls-evaluation} shows, \ToolName can drastically reduce the number of \syscalls available to a userspace application.
For our \SIx{18} tested applications, Nginx and httpd block the least number of \syscalls with \SIx{106} being allowed.
However, without \ToolName, all \SIx{349} \syscalls that Linux 5.0 provides would be available~\cite{LinuxSyscallTable}.
While \ToolName drastically reduces the attack surface, both \CompilerName and \BinName often allow more \syscalls than necessary.
We estimate our approaches' overapproximation to determine how tight the \syscall filters are that an automated approach can generate.
As such, we are the first to demonstrate this for an automated seccomp filter generation tool.

\paragrabf{Setup}
To evaluate our static components' overapproximation, we leverage the functionality of \DynamicName in \LibraryName.
This has the advantage over \textit{strace} that we do not include \syscalls that are needed for setting up the application, \ie we only log \syscalls after the main entry point.
Using this setting, we then either execute the applications test suite or execute the program with different arguments to trigger different code paths, \ie try to trigger as many of the existing \syscalls as possible.
Note that the accuracy of our results depends on the code coverage of the respective test suites.
As was the case in \cref{sec:correctness}, we again argue that despite this not being an exhaustive test, test suites typically cover at least the core functionality of the tested applications and its critical components.
We again substantiate this claim with the code coverage metrics for selected applications shown in \cref{tbl:coverage}.
As the coverage metrics indicate, large parts of the respective applications are executed.
This demonstrates that this is an adequate but not perfect approach to detect the overapproximation of \ToolName.
Furthermore, we are the first to provide an insight into the tightness of automatically generated filters.

Using the aforementioned approach, we obtain a list of \syscalls that the evaluated program issued.
We calculate that list's intersection with the allowed \syscalls as detected by \CompilerName or \BinName.
This gives us a list of \syscalls that our approach allows but that are never executed by the application in our tests.
If a \syscall was triggered that our static approaches block, \DynamicName automatically refines the application's filter list.

\paragrabf{Compiler}
For git, FFmpeg, memcached, redis, sqlite3, httpd, and Nginx, \BinName allows on average \SIx{2} \syscalls more than \CompilerName.
\CompilerName outperforms it due to two points: the heuristics based on function signatures that we apply in the compiler to infer potential indirect call targets, and as it only includes \syscalls that are reachable from \textit{main} and \textit{exit}.
While \BinName has to rely on heuristics as well, the amount of available information is smaller as it needs to disassemble generated code and infer the call graph from it using angr instead of doing it on the source code level where more meta-information is available.
If angr incorrectly infers a call target, it potentially merges \syscalls that are not necessary into a function that is later on used by the target application, which results in overapproximation.

As \cref{tbl:syscalls-evaluation} shows, overapproximation varies between different applications.
Out of the shown busybox utilities, we observed the largest overapproximation for \textit{env}, where only \SI{20}{\percent} of the detected \syscalls are actually used.
For the larger applications, we observe the largest overapproximation in mutool, with only \SI{26.23}{\percent} being used.

\paragrabf{Binary}
For the evaluation of \BinName, we slightly deviate from the outlined setup just to ease the evaluation.
The only difference is that we do not rely on the linked \LibraryName library, but instead use the standalone implementation of \DynamicName.
Hence, we observe a larger amount of \syscalls as we also record \syscalls executed during program startup, similar to \textit{strace}.

In busybox, we overapproximate the most in the \textit{true} utility, where only \SI{37.5}{\percent} are being used.
In the larger applications, we observe the lowest percentage of actually used \syscalls in mutool, with only \SI{21.74}{\percent} being used.

As was the case with the functional-correctness evaluation, we would like to investigate the possibility of better estimating the overapproximation using a coverage-guided fuzzer.
Unfortunately, this is out-of-scope for this paper.

\subsubsection{Mitigating Real-World Exploits}\label{sec:eval-security}
For evaluating the effectiveness of \ToolName in mitigating real-world exploits, we assume an attacker that can either inject shellcode or mount a code-reuse attack~\cite{Szekeres2013sok} in one of our target applications.
We define an exploit as successful if the attacker can exploit a kernel bug from the application context.
These bugs in the kernel either trigger a privilege escalation or result in a denial of service.
As seccomp filters restrict the available \syscalls, they reduce the attack surface of the kernel.

\begin{table}[t]
  \centering
  \begin{tabular}{lc}
    \textbf{\Syscall} & \textbf{Equivalents} \\
    \toprule
     munlockall & munlock \\
     listxattr & llistxattr, flistxattr \\
     epoll\_create & epoll\_create1 \\
     mlockall & mlock, mlock2 \\
     execve & execveat \\
     recvfrom & recvmsg, recvmmsg \\
     writev & pwritev \\
     mknod & mknodat \\
     open & openat \\
     accept & accept4 \\
     getdents & getdents64 \\
     sendto & sendmmsg, sendmsg \\
     getxattr & fgetxattr, lgetxattr \\
     rename & renameat, rename2 \\
     epoll\_ctl & epoll\_ctl\_old \\
    \bottomrule
  \end{tabular}
  \caption{\Syscalls and their equivalents.}
  \label{tbl:alternative-syscalls}
\end{table}

For the evaluation, we extract a list of CVEs from the mitre database~\cite{Mitre} that exploit \syscalls on the x86\_64 Linux kernel.
This results in a total of \SIx{175} CVEs since \SIx{2003}.
From this list, we extract the necessary \syscalls and map them to the corresponding \syscall numbers, resulting in a list of \SIx{231} malicious samples.
The reason for the higher number is that a CVE can be triggered by different \syscalls that are independent of each other.
As some \syscalls have equivalent versions that perform the same action, we extend our list of samples to \SIx{320} by substituting the \syscall numbers where applicable.
We provide a list of these equivalent \syscalls in \cref{tbl:alternative-syscalls}.
As we want to show that \ToolName-sandboxed applications impede the exploitation of unpatched kernel vulnerabilities, we assume a kernel that is vulnerable to all these CVEs.

To determine the effectiveness of \ToolName, we cross-reference the \syscall numbers from each sample with the ones prohibited by our various test applications in \cref{tbl:syscalls-evaluation}.
If one of the \syscalls required for the exploit is not allowed by the application, we determine that this application cannot trigger the exploit in the kernel, indicating that \ToolName increased the security of the system.
We consider both the number of CVEs that we fully mitigate and the number of subvariants mitigated by \ToolName.

\paragrabf{Compiler}
With \CompilerName, we can mitigate around \SI{84.04}{\percent} of the CVEs completely and around \SI{89.42}{\percent} of the subvariants in the case of busybox.
The reason for that is that the busybox utilities are rather small, allowing only a few \syscalls.
With the larger applications, our compiler performs worse with mitigating around \SI{38.08}{\percent} of CVEs completely and around \SI{56.53}{\percent} of the subvariants.
Even though \CompilerName does not perform as well for the larger binaries, it still increases the system's security. %

\paragrabf{Binary}
In busybox, \BinName mitigates around \SI{81.8}{\percent} of the CVEs completely and around \SI{87.9}{\percent} of the subvariants.
In the larger binaries, \BinName can fully mitigate \SI{36.38}{\percent} of the CVEs and \SI{52}{\percent} of the subvariants.

\subsubsection{Preventing Malicious SGX Enclaves}\label{sec:eval:sec:sgx}
Intel SGX enclaves cannot directly execute any \syscalls, but only use functionality provided by the host application.
The host application can use \syscalls to provide this functionality to the enclave.
Schwarz~\etal\cite{Schwarz2019SGXMalware} presented a technique to execute arbitrary \syscalls from an SGX enclave by mounting a ROP attack on the host application.
This allows malicious or exploited enclaves to mount attacks on the kernel.

Weiser~\etal\cite{Weiser2019SGXJail} presented SGXJail as a generic countermeasure for malicious enclaves, preventing them from executing arbitrary \syscalls.
\BinName achieves a similar goal without affecting the performance of required \syscalls.
For the evaluation, we used the public proof-of-concept exploit provided by Schwarz~\etal\cite{Schwarz2019SGXMalware}.
The Intel SGX SDK currently does not support LLVM; hence, we can only evaluate \BinName.
As enclaves cannot contain \syscalls, \BinName only has to scan the host application and allow only \syscalls legitimately used by the host application.
Out of the 349 \syscalls provided by Linux 5.0, 279 (\SI{79.9}{\percent}) are blocked, including \texttt{exec}.
We verified that the benign functionality of the host and enclave is not impacted.
As a result, the malicious (or exploited) enclave cannot run arbitrary programs anymore, and the attack surface is drastically reduced.

\subsection{Comparison to Other Approaches}\label{sec:compare}
Recently, the field of automating seccomp filter generation has gained traction with the publication of two works~\cite{Ghavamnia2020,DeMarinis2020}.
Note that these works were only published after the start of our 6-month long-term case study of \ToolName on Nginx.
In this section, we want to take a closer look at these two approaches and discuss the differences between them and our work.

\paragrabf{Temporal \Syscall Specialization}
Ghavamnia~\etal\cite{Ghavamnia2020} propose an automated approach to detect the used \syscalls during compilation.
Contrary to our approach, they require a multitude of tools for the compilation and link-time optimization that is not supported by every application.
Their approach is limited to applications that can be split into an initialization and serving phase, such as server applications. 
The basic idea is to detect \syscalls used after the server's initialization phase, \ie the point in time where it starts handling requests.
Thus, this approach is not directly applicable to applications that cannot be easily split into an initialization and serving phase, potentially enabling attacks through browsers, malicious PDFs~\cite{Esparza2012,Esparza2014}, messengers~\cite{Silvanovich2020,Gross2020}, and office applications~\cite{Mueller2020,Infuer2019}.
We explicitly consider such applications in our approach (\cf \cref{sec:threat-model}).
Similar to \ToolName, they also extract a sufficiently precise call graph to be able to extract which \syscalls are reachable by the application.
Their approach relies on Andersen's points-to analysis, which is known to not scale with program size~\cite{Andersen1994points,Hind2001pointer}.
We evaluated an orthogonal \textit{has address taken} approach as is used by LLVM's CFI implementation.
As this is already used for the CFI implementation of LLVM, we know that the resulting CFG is reasonably precise as otherwise applications that rely on software-based CFI would not work.
Our evaluation showed that this approach achieves similar results in terms of detected \syscalls as the more complex and slower approach used by Ghavamnia~\etal\cite{Ghavamnia2020}.
As neither Andersen's points-to nor our address taken approach can guarantee a complete CFG, we rely on the more practical address-taken algorithm.
This choice significantly reduces the compile time.
For instance, \syscall extraction for Nginx using Andersen's algorithm shows an increase in compilation time from \SI{1}{\minute} to \SI{83}{\minute} (+\SI{8300}{\percent})~\cite{Ghavamnia2020} compared to an increase of \SI{7.9}{\second} (+\SI{10.53}{\percent}) with \ToolName.
Another difference is that our approach allows for a larger threat model as we also include a potential local attacker instead of just a remote one.

In summary, we have significantly improved the approach's performance while maintaining accuracy and security properties.
Additionally, we allow for the approach to be applicable to a broader range of applications, including local applications that are commonly exploited.
We also provide an evaluation of the tightness of the resulting filters.

\paragrabf{Sysfilter}
A second approach, \textit{sysfilter}~\cite{DeMarinis2020} focuses on extracting \syscalls from existing binaries.
While sysfilter and \BinName share the same goal, the approaches differ in the used tools, \ie \BinName relies on the angr framework that already supports parts of what sysfilter manually implemented.
Both approaches show similar success rates in mitigating exploits in their respective test sets.

Sysfilter provides no analysis of the approach's overapproximation, making it hard to estimate how tight the resulting \syscall filters are.
Hence, we perform such an analysis to show differences between the approaches.
As we discussed in \cref{sec:correctness,sec:eval-overapproximation}, obtaining a ground truth is infeasible and would require computational intensive formal proofs.
Hence, we need another source for a reliable baseline to which we can compare the results of the evaluation for \BinName and sysfilter.

To provide this baseline, we rely on the results of \CompilerName when generating a static binary.
The reason why we do this is twofold.
First, the compiler has the most information about the application as it needs to generate a functioning binary, \ie it needs to know which functions are actually required and called.
The second reason is based on what a compiler like clang does when it generates the static binary that we use.
When generating this binary, the compiler already removes all unnecessary functions, \ie functions that are never called and never have their address taken, from the binary.
So the resulting binary only contains functions and their respective \syscalls if the compiler determined a potential path to the respective function.
Therefore, any \syscall found by the two binary tools within the static binary can be reached and is necessary for the application to work correctly.
This number may differ from the one detected by \CompilerName due to the inherent overapproximation of the function signature heuristic, \ie read and write have the same function signature, so if one is used, the other one is automatically included in the set.
In this case, the \syscall of a function is included even though the compiler removed the function's actual code.
Nevertheless, we expect the numbers to be in a similar range.

In this evaluation, both sysfilter and \BinName work on the exact same static binaries.
We ensured that the binary still contains the stack unwinding information (\textit{.eh\_frame}) and other necessary sections (\textit{.init, .fini}) on which sysfilter relies for its precise disassembly.
While sysfilter notes that one requirement is a position-independent binary, we note that there is no reason for such a requirement as additional tasks that sysfilter performs for PIC binaries, \ie relocations or checking the dynamic symbol table, are by the design of static binaries simply not necessary.
Building the call graph does not depend on these steps either.
In fact, for binary analysis tools like sysfilter and \BinName, a static binary can be considered the most straightforward use case as all information is already contained within the single binary.

In the evaluation, we consider two different modes of sysfilter, \ie the default behavior that prunes the call graph based on a reachability analysis and the universal approach that assumes that every function is reachable by every other function.
As the binary is compiled statically, we expect that both modes produce the same result as only functions that are reachable from the main entry point are included.
We show the result of this analysis in \cref{tbl:sysfilter-comparison}.

\begin{table}[t]
  \centering
  \resizebox{\hsize}{!}{
    \begin{tabular}{lcccc}
      \textbf{Binary} & \CompilerName & \BinName  & \makecell{sysfilter             \\ (vacuumed-fcg)} & \makecell{sysfilter\\ (universal-fcg)} \\
      \toprule
      FFmpeg          & \SIx{63}      & \SIx{53}  & \SIx{18}            & \SIx{53}  \\
      busybox         & \SIx{163}     & \SIx{144} & \SIx{15}            & \SIx{152} \\
      Redis-server    & \SIx{85}      & \SIx{74}  & \SIx{12}            & \SIx{74}  \\
      \bottomrule
    \end{tabular}
  }
  \caption{The number of extracted \syscalls by \CompilerName, \BinName, and the two modes of sysfilter.}
  \label{tbl:sysfilter-comparison}
\end{table}

As our analysis shows, the assumption that both modes of sysfilter produce the same result does not hold as the pruning-based mode significantly underapproximates in all three evaluated binaries.
The low number of detected \syscalls hints at some mistake in the pruning algorithm as the number is too low for such complex applications.
In two out of three binaries, \BinName and sysfilter using the universal approach produce the exact same result while the third binary only shows a small difference of \SIx{8} \syscalls.
In this case, the difference to \CompilerName is within an expected range due to the overapproximation of \CompilerName.
This is not true for the pruning-based approach of sysfilter as the difference is too large, and the number of detected \syscalls is lower than the number of \syscalls that are actually used (\cf \cref{tbl:syscalls-evaluation}).
Interestingly, the universal-fcg implementation of sysfilter also supports our observation that a PIC binary is not a requirement for these types of binary analysis tools as it produces similar results to \BinName, contradicting the statement by its developers.
Nevertheless, there is still a difference in the operation between the universal-fcg approach of sysfilter and \BinName as the latter achieves this result by not assuming that every function is reachable by every other function.
Instead, it still builds a correct call graph and derives the information from it, which fails for the vacuum-fcg approach of sysfilter.

We further investigated why the number of found \syscalls is so low for the pruning-based approach of sysfilter.
This analysis showed that during the pruning, the main function is removed from the set of reachable functions, which results in the whole application being removed from the analysis.
We leave the analysis of whether this is purely an implementation fault or a hint that this is a general fault in the approach for future work as this is out of scope for this paper.

\section{Discussion}\label{sec:discussion}
\paragraph{Limitations and Future Work.}
One of the limitations is the performance of seccomp as it is quite slow~\cite{Hromatka2018,Tizen2014}. %
Unfortunately, this is a limitation imposed by the underlying system and not a weakness of \ToolName itself.
Improving the performance of seccomp is considered out of scope for this paper, and there is no suitable alternative.

Another limitation is the overapproximation of both the compiler- and binary-based approaches.
Fast and reliable points-to analysis with limited overapproximation is still an unsolved problem as previous work has shown~\cite{Hind2001pointer}.
In some cases, we also exhibit the opposite effect in \textit{angr} that it is not able to detect the call target of an indirect call, hence missing a potentially reachable \syscall.

In future work, we will investigate the possibility of extending the \syscall filtering with argument tracking.
While detecting the constant arguments from a \syscall is possible, the problem is the propagation of the information throughout the call graph so that in the end, only the argument remains that is needed.
If we can solve this problem, we can restrict \syscalls even further.
For instance, for the \texttt{exec} \syscalls, we can detect hardcoded paths and install filters that only allow it with this path.
For \texttt{mprotect}, we can then further restrict the possible set of permissions so that only those are allowed that are used, potentially preventing \texttt{mprotect} with executable permissions.
Furthermore, we will investigate performance improvements for seccomp as well as alternatives.
We also want to investigate the possibility of using coverage-guided fuzzers to estimate the overapproximation of automated seccomp filter generation tools.

\paragraph{Related Work.}
For related work, we mostly focus on work that tried to automate a sandboxing process as this is also what was done in this work.
Previous work has already focused on reducing the attack surface of an application by removing unused code from an application.
One of the first approaches for library debloating was proposed by Mulliner and Neugschwandtner~\cite{mulliner2015breaking} based on removing non-imported functions from a shared library during load time.
This approach has been further improved by Quach~\etal\cite{quach2018debloating} by removing all unused functions from shared libraries during load time by extending the compiler and the loader.
Agadakos~\etal\cite{agadakos2019nibbler} proposed a binary-level approach for library debloating.
It is based on function boundary detection and dependency identification to identify and erase unused functions.
Davidson~\etal\cite{Davidsson2019} performed an analysis of the complete software stack for web applications to create specialized libraries based on the requirements for PHP code and the server binaries.
Mishra and Polychronakis~\cite{mishra2018shredder} proposed \textit{Shredder}, which instruments binaries to restrict arguments to critical system APIs to a predetermined allowlist.
Another approach is to apply data dependency analysis for fine-grained library customization of static libraries~\cite{song2018fine}.
Ghavamnia~\etal\cite{Ghavamnia2020} propose a similar approach to \CompilerName, but at the cost of a much higher analysis time during compilation but with similar accuracy in detecting \syscalls.
Wagner and Dean~\cite{wagner2000} propose a static approach to build an IDS that uses a similar approach to \CompilerName for pointer analysis to extract a model of expected application behavior.
In general, several papers have proposed static analysis of \syscalls for anomaly detection and IDS~\cite{Feng2004}.
Rajagopalan~\etal\cite{Rajagopalan2005} propose to replace \syscalls with authenticated \syscalls that specify a policy and provide a cryptographic message authentication code that guarantees the integrity of the \syscall.

Other approaches propose to reduce the attack surface by relying on training to identify the unused code sections.
Ghaffarinia and Hamlen~\cite{ghaffarinia2019binary} rely on training to limit control flow transfers to not authorized sections of code.
An approach without access to the source code uses training and heuristics to identify and remove unnecessary basic blocks~\cite{qian2019razor}.

Previous work focused mostly on C/C++ software with few solutions for software in other languages.
For Java, one approach uses static code analysis to remove unused classes and methods~\cite{jian2016jred}.
For PHP, Azad~\etal\cite{azad2019debloating} proposed a framework using dynamic analysis to remove superfluous features.

\section{Conclusion}\label{sec:conclusion}

\ToolName is an automated approach for sandboxing native Linux userspace applications on the \syscall level.
It mainly relies on static analysis to identify \syscalls required by an application and blocks the unused \syscalls.
\ToolName also supports an optional dynamic refinement phase that can be used to restrict the \syscall filters further.
In contrast to existing solutions, \ToolName has fewer requirements and limitations.
For instance, the compiler-based approach is up to factor \SIx{73} faster than previous work without any loss in accuracy.
For the binary-level analysis, we lifted the requirement of a position-independent binary, making our approach applicable to a broader set of applications.
We demonstrated that in a selection of 18 real-world applications, \ToolName can, on average, block \SIx{302} \syscalls (\SI{86.5}{\percent}) when used on the source level, and \SIx{288} (\SI{82.5}{\percent}) when used on the binary level.
Our analysis showed that the compiler- and binary-based approach prevent exploitation of more than \SI{63}{\percent} and \SI{61}{\percent}, respectively, of Linux kernel CVEs, which can be triggered via \syscalls, in case the selected applications have been exploited.
Moreover, \ToolName can reduce the attack surface by blocking the dangerous \texttt{exec} \syscall in the majority of tested applications.
We have provided insights into the functional correctness and the overapproximation of \ToolName, which we substantiated by code coverage metrics of the respective tests.
Additionally, we showed the results of a 6month long-term evaluation of Nginx on a real server.
Finally, our work shows that automated sandboxing is feasible and increases platform security without manual effort.

\ifAnon

\else
\section*{Acknowledgments}
This work has been supported by the Austrian Research Promotion Agency (FFG) via the project ESPRESSO, which is funded by the province of Styria and the Business Promotion Agencies of Styria and Carinthia.
This project has received funding from the European Research Council (ERC) under the European Union's Horizon 2020 research and innovation program (grant agreement No 681402).
Additional funding was provided by generous gifts from ARM, Intel, and Red Hat.
Any opinions, findings, and conclusions or recommendations expressed in this paper are those of the authors and do not necessarily reflect the views of the funding parties.
\fi

{
\footnotesize
\bibliographystyle{IEEEtranS}
\bibliography{main}
}

\end{document}

%% file: images/overview.tikz
\tikzstyle{dbox} += [draw,thick]
\tikzstyle{phase} += [font=\bfseries\large]

\begin{tikzpicture}

\begin{scope}[shift={(0,0)}]
  \draw[dbox] (-0.5,-3) rectangle node[yshift=2.5cm,phase] {P1: Static Analysis} +(7.5,6);
  \node (src) at (0.5,1) [minimum width=1cm,minimum height=1.25cm] {\parbox{1cm}{\centering \includegraphics[width=1cm]{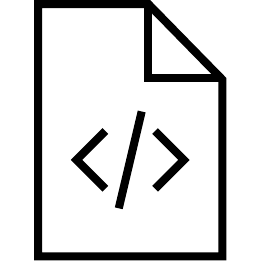}\\ Source}};
  \node (bin) at (0.5,-2) [minimum width=1cm,minimum height=1.25cm] {\parbox{1cm}{\centering \includegraphics[width=1cm]{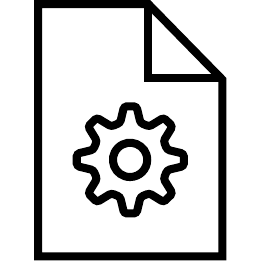}\\ Binary}};
  \node (binfilter) at (6,-.5) [minimum width=1cm,minimum height=1.25cm] {\parbox{2cm}{\centering \includegraphics[width=1cm]{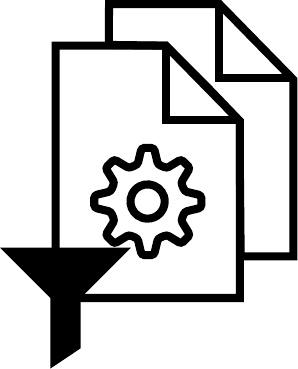}\\ Annotated\\Binary File(s)}};
  \node (llvm) at (3,1) [draw,minimum width=2cm,minimum height=1cm] {\parbox{2cm}{\centering Source\\Analyzer}};
  \node (binan) at (3,-2) [draw,minimum width=2cm,minimum height=1cm] {\parbox{2cm}{\centering Binary\\Analyzer}};
\end{scope}

\begin{scope}[shift={(10,0)}]
  \draw[dbox,fill=gray!30] (-1.8,-3) rectangle node[yshift=2.5cm,phase] {P2: Dynamic Refinement} +(5.8,6);
  \node (dynan) at (0,-.5) [draw,minimum width=2cm,minimum height=1cm] {\parbox{2cm}{\centering Dynamic\\Analyzer}};
  \node (binfilter_refined) at (3,-.5) [minimum width=1cm,minimum height=1.25cm] {\parbox{2cm}{\centering \includegraphics[width=1cm]{images/binfilters}\\ Annotated\\Binary File(s)}};
\end{scope}

\begin{scope}[shift={(16.5,0)}]
  \node (wrapper) at (0,-2) [draw,minimum width=2cm,minimum height=1cm] {\parbox{2cm}{\centering Chestnut\\Generator}};
  \node (rewriter) at (0,1) [draw,minimum width=2cm,minimum height=1cm] {\parbox{2cm}{\centering Chestnut\\Patcher}};
  \node (binrewrite) at (3,1) [minimum width=0.7cm,minimum height=1.25cm] {\parbox{1.1cm}{\centering \includegraphics[width=1cm]{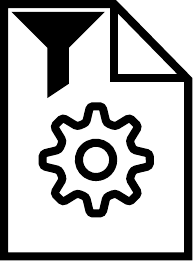}\\ Sandboxed\\Binary}};
  \node (binsandbox) at (3,-2) [minimum width=0.7cm,minimum height=1.25cm] {\parbox{1.1cm}{\centering \includegraphics[width=1cm]{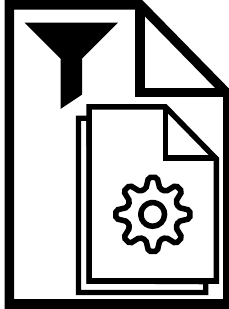}\\ Wrapped\\Binary}};
\end{scope}

\begin{scope}
  \draw[->,>=stealth,out=0,in=180,thick] (src.east) to (llvm.west);
  \draw[->,>=stealth,out=0,in=180,thick] (bin.east) to (binan.west);
  \draw[->,>=stealth,out=0,in=180,thick] (binan.east) to (binfilter.west);
  \draw[->,>=stealth,out=0,in=180,thick] (llvm.east) to (binfilter.west);
\end{scope}

\begin{scope}
  \draw[->,>=stealth,out=0,in=180,thick] (binfilter.east) to (dynan.west);
  \draw[->,>=stealth,out=0,in=180,thick] (dynan.east) to (binfilter_refined.west);
  \draw[->,>=stealth,out=0,in=180,thick] (binfilter_refined.east) to (rewriter.west);
  \draw[->,>=stealth,out=0,in=180,thick] (binfilter_refined.east) to (wrapper.west) node[above,yshift=1.3cm] {or};
\end{scope}

\begin{scope}
  \draw[->,>=stealth,out=0,in=180,thick] (wrapper.east) to (binsandbox.west);
  \draw[->,>=stealth,out=0,in=180,thick] (rewriter.east) to (binrewrite.west);
\end{scope}

\end{tikzpicture}

%% file: images/compiler-flow.tikz
\tikzset{multiple/.style = {fill=white,draw=black,thick,minimum height = 1cm,minimum
           width=2cm},
         ordinary/.style = {rectangle,draw,thick,minimum height = 1cm,minimum width=2cm,rounded corners}}
\tikzstyle{arrow} = [thick,->,>=stealth]

\begin{tikzpicture}[scale=0.5]
  \node[] at (0,0) (source) {\parbox{1cm}{\centering \includegraphics[width=1cm]{images/srcfile}\\ Sources}};
  \node[right=0.5cm of source,align=center] (stage1) {\parbox{2cm}{\centering \includegraphics[width=1cm]{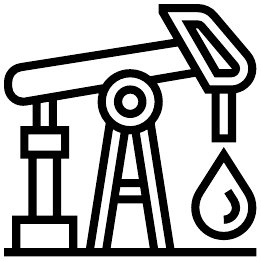}\\ Extract System Call Number}};
  \node[right=0.5cm of stage1,align=center] (stage2) {\parbox{2cm}{\centering \includegraphics[width=1.5cm]{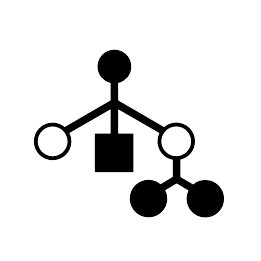}\\ Build Call Graph}};
  \node[below right=-1cm and 0.5cm of stage2,align=center] (annotated) {\parbox{2cm}{\centering \includegraphics[width=1cm]{images/binfilters}\\ Annotated\\Binary File(s)}};
  \node[above right=-1cm and 0.5cm of stage2] (libs) {\parbox{2cm}{\centering \includegraphics[width=1cm]{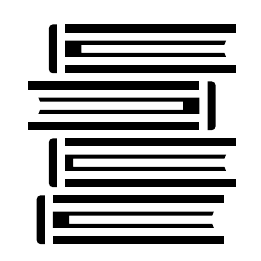}\\ Libraries}};
  \node[right=3cm of stage2,align=center] (stage3) {\parbox{2cm}{\centering \includegraphics[width=1cm]{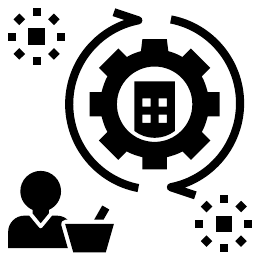}\\ Generate System Call List}};
  \node[right=0.5cm of stage3,align=center] (executable) {\parbox{2cm}{\centering \includegraphics[width=1cm]{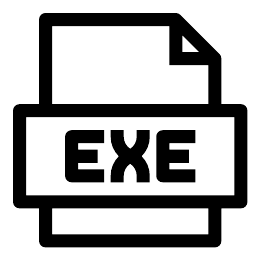}\\ Annotated Executable}};
  \node[right=0.5cm of executable,align=center] (run) {\parbox{2cm}{\centering \includegraphics[width=1cm]{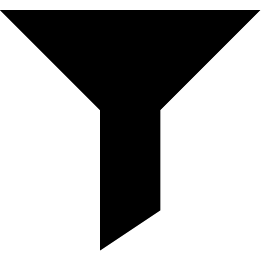}\\ Install Filters}};

  \draw[arrow] (source) to (stage1);
  \draw[arrow] (stage1) to (stage2);
  \draw[arrow,out=0,in=135] (stage2.east) to (annotated.west);
  \draw[arrow,out=0,in=180] (libs.east) to (stage3.west);
  \draw[arrow,out=0,in=180] (annotated.east) to (stage3.west);
  \draw[arrow] (stage3.east) to (executable.west);
  \draw[arrow] (executable.east) to (run.west);  
\end{tikzpicture}

%% file: images/symbolic_backward_execution.tikz
\begin{tikzpicture}

  \node[right] at (0,2.5) {\texttt{mov}};
  \node[right] at (0.8,2.5) {\texttt{\$0x1,}};
  \node[right] at (1.8,2.5) {\texttt{\%bl}};
  
  \node[right] at (0,2) {\texttt{xor}};
  \node[right] at (0.8,2) {\texttt{\%edi,}};
  \node[right] at (1.8,2) {\texttt{\%edi}};
  
  \node[right] at (0,1.5) {\texttt{mov}};
  \node[right] at (0.8,1.5) {\texttt{\%ebx,}};
  \node[right] at (1.8,1.5) {\texttt{\%eax }};
  \node[right] at (0,1) {\texttt{lea}};
  \node[right] at (0.8,1) {\texttt{0xf(\%rip),}};
  \node[right] at (2.8,1) {\texttt{\%rsi}};
  \node[right] at (0,0.5) {\texttt{mov}};
  \node[right] at (0.8,0.5) {\texttt{\$0xd,}};
  \node[right] at (1.8,0.5) {\texttt{\%edx}};
  \node[right] at (0,0) {\texttt{syscall }};

  \draw[->,>=stealth,thick] (4,0) to (4,2.5);
  
  \node[right] at (4.5,0) {\texttt{rax} = ?};
  \node[right] at (4.5,0.5) {\texttt{rax} = ?};
  \node[right] at (4.5,1) {\texttt{rax} = ?};
  \node[right] at (4.5,1.5) {\texttt{rax} = \texttt{rbx} = ?};
  \node[right] at (4.5,2) {\texttt{rax} = \texttt{rbx} = ?};
  \node[right] at (4.5,2.5) {\texttt{rax} = \texttt{rbx} = \$0x1};

  \draw[red!60!black,thick] (0.75,0) ellipse (0.8cm and 0.25cm);
  \draw[red!60!black,thick] (2.25,1.5) ellipse (0.5cm and 0.25cm);
  \draw[red!60!black,thick] (1.25,1.5) ellipse (0.5cm and 0.25cm);
  \draw[red!60!black,thick] (2.25,2.5) ellipse (0.5cm and 0.25cm);
  \draw[red!60!black,thick] (1.25,2.5) ellipse (0.5cm and 0.25cm);
  
  \draw[red!60!black,thick,out=90,in=90,->,>=stealth] (2.25,2.75) to (1.25,2.75);
  \draw[red!60!black,thick,out=270,in=270,->,>=stealth] (2.25,1.25) to (1.25,1.25);
  \draw[red!60!black,thick,out=90,in=270,->,>=stealth] (1.25,1.75) to (2.25,2.25);
  \draw[red!60!black,thick,out=90,in=270,->,>=stealth] (0.85,0.25) to (2.45,1.26);
\end{tikzpicture}

%% file: images/syscall_mapping.tikz
\begin{tikzpicture}[yscale=0.75]
\usetikzlibrary{shapes}

\node (puts) [draw=black,ellipse,minimum width=2cm] at (0,2) {puts};
\node (fork) [draw=black,ellipse,minimum width=2cm] at (0,1) {fork};
\node (system) [draw=black,ellipse,minimum width=2cm] at (7,1.5) {system};

\node (writev) [draw=black,minimum width=2.75cm,minimum height=0.4cm] at (3.5,2) {sys\_writev};
\node (clone) [draw=black,minimum width=2.75cm,minimum height=0.4cm] at (3.5,1.25) {sys\_clone};
\node (futex) [draw=black,minimum width=2.75cm,minimum height=0.4cm] at (3.5,0.5) {sys\_futex};
\node (wait) [draw=black,minimum width=2.75cm,minimum height=0.4cm] at (3.5,-0.25) {sys\_wait4};
\node (execve) [draw=black,minimum width=2.75cm,minimum height=0.4cm] at (3.5,-1) {sys\_execve};
\node (mask) [draw=black,minimum width=2.75cm,minimum height=0.4cm] at (3.5,-1.75) {sys\_rt\_sigprocmask};

\draw[->,>=stealth,in=180,out=0] (puts.east) to (writev.west);
\draw[->,>=stealth,in=180,out=0] (fork.east) to (writev.west);
\draw[->,>=stealth,in=180,out=0] (fork.east) to (clone.west);
\draw[->,>=stealth,in=180,out=0] (fork.east) to (futex.west);
\draw[->,>=stealth,in=0,out=180,looseness=0.6] (system.west) to (clone.east);
\draw[->,>=stealth,in=0,out=180,looseness=0.6] (system.west) to (wait.east);
\draw[->,>=stealth,in=0,out=180,looseness=0.6] (system.west) to (execve.east);
\draw[->,>=stealth,in=0,out=180,looseness=0.6] (system.west) to (mask.east);
\draw[->,>=stealth,in=0,out=180,looseness=0.6] (system.west) to (writev.east);

\end{tikzpicture}

%% file: images/dynamic-trace.tikz
\tikzstyle{startstop} = [rectangle, rounded corners, minimum width=3cm, minimum height=0.75cm,text centered, draw=black,fill=black!2]
\tikzstyle{kernel} = [rectangle, minimum width=3cm, minimum height=0.75cm, text centered, draw=black,fill=black!2]
\tikzstyle{process} = [rectangle, minimum width=3cm, minimum height=0.75cm, text centered, draw=black, fill=black!10]
\tikzstyle{decision} = [diamond, minimum width=3cm, minimum height=0.75cm, text centered, draw=black, fill=white,thick]
\tikzstyle{arrow} = [thick,->,>=stealth]
\usetikzlibrary{shapes.geometric, arrows}

\begin{tikzpicture}[yscale=.55]

\draw[kernel] (7,-1) rectangle +(4,2);
\node[left] at (11,1.55) {\large Kernel};

\draw[arrow] (6,-1.75) -| (8,-1) node[midway,below] {System Call};
\draw[arrow] (8,-1) |- (6,1.5) node[midway,above,xshift=-1cm] {notify};
\draw[draw=red!60!black] (8.5,-0.5) rectangle +(2,1) node[midway] {System Call};

\draw[startstop] (0,0) rectangle +(6,3);
\node[right] at (0.5,1.5) {\Large Tracer};
\draw[arrow] (4.5,2.95) -- (4.5,3.75) -| node[midway,below,xshift=-1cm] {allow} (9.5,0.5);
\draw[arrow] (9.5,-0.5) |- node[midway,right,yshift=0.5cm] {return} (6,-2.75);

\node[decision] at (4.5,1.5) {\parbox{1cm}{\centering Log Syscall}};
\draw[process] (0,-3) rectangle +(6,1.5) node [midway,xshift=-1cm] {\Large Tracee};

\draw[arrow] (0.5,0) -- (0.5,-1.5) node[midway,right] {Trace};

\end{tikzpicture}

%% file: main.bbl
\begin{thebibliography}{10}
\providecommand{\url}[1]{#1}
\csname url@samestyle\endcsname
\providecommand{\newblock}{\relax}
\providecommand{\bibinfo}[2]{#2}
\providecommand{\BIBentrySTDinterwordspacing}{\spaceskip=0pt\relax}
\providecommand{\BIBentryALTinterwordstretchfactor}{4}
\providecommand{\BIBentryALTinterwordspacing}{\spaceskip=\fontdimen2\font plus
\BIBentryALTinterwordstretchfactor\fontdimen3\font minus
  \fontdimen4\font\relax}
\providecommand{\BIBforeignlanguage}[2]{{%
\expandafter\ifx\csname l@#1\endcsname\relax
\typeout{** WARNING: IEEEtranS.bst: No hyphenation pattern has been}%
\typeout{** loaded for the language `#1'. Using the pattern for}%
\typeout{** the default language instead.}%
\else
\language=\csname l@#1\endcsname
\fi
#2}}
\providecommand{\BIBdecl}{\relax}
\BIBdecl

\bibitem{agadakos2019nibbler}
I.~Agadakos, D.~Jin, D.~Williams-King, V.~P. Kemerlis, and G.~Portokalidis,
  ``Nibbler: debloating binary shared libraries,'' in \emph{ACSAC}, 2019.

\bibitem{Andersen1994points}
L.~O. Andersen, ``{Program Analysis and Specialization for the C Programming
  Language},'' Ph.D. dissertation, May 1994.

\bibitem{azad2019debloating}
B.~A. Azad, P.~Laperdrix, and N.~Nikiforakis, ``Less is more: Quantifying the
  security benefits of debloating web applications,'' in \emph{{USENIX}
  Security Symposium}, August 2019.

\bibitem{Bletsch2011JOP}
T.~K. Bletsch, X.~Jiang, V.~W. Freeh, and Z.~Liang, ``{Jump-oriented
  programming: a new class of code-reuse attack},'' in \emph{{AsiaCCS}}, 2011.

\bibitem{Bosman2014SROP}
E.~Bosman and H.~Bos, ``Framing signals - {A} return to portable shellcode,''
  in \emph{{S\&P}}, 2014.

\bibitem{Bovet2013elf}
\BIBentryALTinterwordspacing
D.~P. Bovet, ``{Special sections in Linux binaries},'' January 2013. [Online].
  Available: \url{https://lwn.net/Articles/531148/}
\BIBentrySTDinterwordspacing

\bibitem{Canella2019A}
C.~Canella, J.~Van~Bulck, M.~Schwarz, M.~Lipp, B.~von Berg, P.~Ortner,
  F.~Piessens, D.~Evtyushkin, and D.~Gruss, ``{A Systematic Evaluation of
  Transient Execution Attacks and Defenses},'' in \emph{USENIX Security
  Symposium}, 2019, extended classification tree and PoCs at
  https://transient.fail/.

\bibitem{Carlini2014COP}
N.~Carlini and D.~A. Wagner, ``{ROP} is still dangerous: Breaking modern
  defenses,'' in \emph{{USENIX} Security Symposium}, 2014.

\bibitem{Checkoway2010JOP}
S.~Checkoway, L.~Davi, A.~Dmitrienko, A.~Sadeghi, H.~Shacham, and M.~Winandy,
  ``Return-oriented programming without returns,'' in \emph{{CCS}}, 2010.

\bibitem{Chen2005data}
S.~Chen, J.~Xu, E.~C. Sezer, P.~Gauriar, and R.~K. Iyer, ``Non-control-data
  attacks are realistic threats.'' in \emph{USENIX Security Symposium}, 2005.

\bibitem{ChromiumSeccomp}
\BIBentryALTinterwordspacing
Chromium, ``{Linux Sandboxing}.'' [Online]. Available:
  \url{https://chromium.googlesource.com/chromium/src/+/0e94f26e8/docs/linux_sandboxing.md}
\BIBentrySTDinterwordspacing

\bibitem{Corbet2020syscall}
J.~Corbet, ``{Constant-action bitmaps for seccomp()},'' 2020.

\bibitem{Davidsson2019}
N.~Davidsson, A.~Pawlowski, and T.~Holz, ``Towards automated
  application-specific software stacks,'' in \emph{{ESORICS}}, 2019.

\bibitem{DeMarinis2020}
N.~DeMarinis, K.~Williams-King, D.~Jin, R.~Fonseca, and V.~P. Kemerlis,
  ``{sysfilter: Automated System Call Filtering for Commodity Software},'' in
  \emph{RAID}, October 2020.

\bibitem{Drysdale2014syscall}
\BIBentryALTinterwordspacing
D.~Drysdale, ``{Anatomy of a system call, part 2},'' 2014. [Online]. Available:
  \url{https://lwn.net/Articles/604515/}
\BIBentrySTDinterwordspacing

\bibitem{Drysdale2015elf}
\BIBentryALTinterwordspacing
------, ``{How programs get run: ELF binaries},'' February 2015. [Online].
  Available: \url{https://lwn.net/Articles/631631/}
\BIBentrySTDinterwordspacing

\bibitem{Edge2012libseccomp}
\BIBentryALTinterwordspacing
J.~Edge, ``{A library for seccomp filters},'' April 2012. [Online]. Available:
  \url{https://lwn.net/Articles/494252/}
\BIBentrySTDinterwordspacing

\bibitem{Edge2015seccomp}
\BIBentryALTinterwordspacing
------, ``{A seccomp overview},'' September 2015. [Online]. Available:
  \url{https://lwn.net/Articles/656307/}
\BIBentrySTDinterwordspacing

\bibitem{Esparza2012}
J.~M. Esparza, ``{Static analysis of a CVE-2011-2462 PDF exploit},'' 2012.

\bibitem{Esparza2014}
------, ``{Quick analysis of the CVE-2013-2729 obfuscated exploits},'' 2014.

\bibitem{Feng2004}
H.~H. Feng, J.~T. Giffin, Y.~Huang, S.~Jha, W.~Lee, and B.~P. Miller,
  ``{Formalizing sensitivity in static analysis for intrusion detection},'' in
  \emph{{S\&P}}, 2004.

\bibitem{firejail2016firejail}
\BIBentryALTinterwordspacing
Firejail, ``{Firejail Security Sandbox},'' 2016. [Online]. Available:
  \url{https://firejail.wordpress.com/}
\BIBentrySTDinterwordspacing

\bibitem{ghaffarinia2019binary}
M.~Ghaffarinia and K.~W. Hamlen, ``Binary control-flow trimming,'' in
  \emph{CCS}, 2019.

\bibitem{Ghavamnia2020}
S.~Ghavamnia, T.~Palit, S.~Mishra, and M.~Polychronakis, ``{Temporal System
  Call Specialization for Attack Surface Reduction},'' in \emph{{USENIX
  Security Symposium}}, August 2020.

\bibitem{Goktas2014COP}
E.~G{\"{o}}ktas, E.~Athanasopoulos, H.~Bos, and G.~Portokalidis, ``Out of
  control: Overcoming control-flow integrity,'' in \emph{{S\&P}}, 2014.

\bibitem{Goldberg1996secure}
I.~Goldberg, D.~Wagner, R.~Thomas, E.~A. Brewer \emph{et~al.}, ``A secure
  environment for untrusted helper applications: Confining the wily hacker,''
  in \emph{USENIX Security Symposium}, 1996.

\bibitem{Gross2020}
S.~Gro{\ss}, ``{Remote iPhone Exploitation Part 1: Poking Memory via iMessage
  and CVE-2019-8641},'' 2020.

\bibitem{Hind2001pointer}
M.~Hind, ``Pointer analysis: Haven't we solved this problem yet?'' in
  \emph{PASTE}, 2001.

\bibitem{Holzmann2015}
G.~J. Holzmann, ``Code inflation,'' \emph{IEEE Software}, 2015.

\bibitem{Hromatka2018}
T.~Hromatka, ``{seccomp and libseccomp performance improvements},'' 2018.

\bibitem{Android2017seccomp}
\BIBentryALTinterwordspacing
G.~Inc., ``{Seccomp filter in Android O},'' July 2017. [Online]. Available:
  \url{https://android-developers.googleblog.com/2017/07/seccomp-filter-in-android-o.html}
\BIBentrySTDinterwordspacing

\bibitem{Google2019sandbox2}
\BIBentryALTinterwordspacing
------, ``{Sandbox2},'' 2019. [Online]. Available:
  \url{https://developers.google.com/sandboxed-api/docs/sandbox2/overview}
\BIBentrySTDinterwordspacing

\bibitem{Infuer2019}
A.~Inf{\"u}hr, ``{Libreoffice (CVE-2018-16858) - Remote Code Execution via
  Macro/Event execution},'' February 2019.

\bibitem{jian2016jred}
Y.~Jiang, D.~Wu, and P.~Liu, ``Jred: Program customization and bloatware
  mitigation based on static analysis,'' in \emph{COMPSAC}, 2016.

\bibitem{Jurczyk2015gadget}
M.~Jurczyk and G.~Coldwind, ``{Permissions overview},'' \emph{Insomni'hack},
  2015.

\bibitem{Kemerlis2015}
V.~Kemerlis, ``Protecting commodity operating systems through strong kernel
  isolation,'' Ph.D. dissertation, Columbia University, 2015.

\bibitem{Kemerlis2014}
V.~P. Kemerlis, M.~Polychronakis, and A.~D. Keromytis, ``ret2dir: Rethinking
  kernel isolation,'' in \emph{USENIX Security Symposium}, 2014.

\bibitem{Kemerlis2012}
V.~P. Kemerlis, G.~Portokalidis, and A.~D. Keromytis, ``kguard: Lightweight
  kernel protection against return-to-user attacks,'' in \emph{{USENIX}
  Security Symposium}, 2012.

\bibitem{Kim2014}
Y.~Kim, R.~Daly, J.~Kim, C.~Fallin, J.~H. Lee, D.~Lee, C.~Wilkerson, K.~Lai,
  and O.~Mutlu, ``{Flipping Bits in Memory Without Accessing Them: An
  Experimental Study of DRAM Disturbance Errors},'' in \emph{ISCA}, 2014.

\bibitem{Kocher2019}
P.~Kocher, J.~Horn, A.~Fogh, D.~Genkin, D.~Gruss, W.~Haas, M.~Hamburg, M.~Lipp,
  S.~Mangard, T.~Prescher, M.~Schwarz, and Y.~Yarom, ``{Spectre Attacks:
  Exploiting Speculative Execution},'' in \emph{S\&P}, 2019.

\bibitem{Lan2015loop}
B.~Lan, Y.~Li, H.~Sun, C.~Su, Y.~Liu, and Q.~Zeng, ``Loop-oriented programming:
  a new code reuse attack to bypass modern defenses,'' in \emph{IEEE
  Trustcom/BigDataSE/ISPA}, 2015.

\bibitem{Lattner2004}
C.~Lattner and V.~S. Adve, ``{LLVM:} {A} compilation framework for lifelong
  program analysis {\&} transformation,'' in \emph{{IEEE} / {ACM} International
  Symposium on Code Generation and Optimization -- {CGO}}, 2004.

\bibitem{Lin2018container}
X.~Lin, L.~Lei, Y.~Wang, J.~Jing, K.~Sun, and Q.~Zhou, ``A measurement study on
  {Linux} container security: Attacks and countermeasures,'' in \emph{ACSAC},
  2018.

\bibitem{LinuxSyscallTable}
\BIBentryALTinterwordspacing
Linux, ``64-bit system call numbers and entry vectors,'' 2019. [Online].
  Available:
  \url{https://github.com/torvalds/linux/blob/master/arch/x86/entry/syscalls/syscall_64.tbl}
\BIBentrySTDinterwordspacing

\bibitem{Lipp2018meltdown}
M.~Lipp, M.~Schwarz, D.~Gruss, T.~Prescher, W.~Haas, A.~Fogh, J.~Horn,
  S.~Mangard, P.~Kocher, D.~Genkin, Y.~Yarom, and M.~Hamburg, ``{Meltdown:
  Reading Kernel Memory from User Space},'' in \emph{USENIX Security
  Symposium}, 2018.

\bibitem{Ma2011directed}
K.-K. Ma, K.~Y. Phang, J.~S. Foster, and M.~Hicks, ``Directed symbolic
  execution,'' in \emph{International Static Analysis Symposium}, 2011.

\bibitem{McCanne1993bsd}
S.~McCanne and V.~Jacobson, ``{The BSD Packet Filter: A New Architecture for
  User-level Packet Capture},'' in \emph{{USENIX} Winter}, January 1993.

\bibitem{McPhee1972integrity}
W.~S. {McPhee}, ``Operating system integrity in os/vs2,'' \emph{IBM Systems
  Journal}, 1974.

\bibitem{Miller2019Bluehat}
M.~Miller, ``{Trends, challenges, and strategic shifts in the software
  vulnerability mitigation landscape},'' \emph{Bluehat IL}, February 2019.

\bibitem{mishra2018shredder}
S.~Mishra and M.~Polychronakis, ``Shredder: Breaking exploits through api
  specialization,'' in \emph{{ACSAC}}, 2018.

\bibitem{Mozilla2016seccomp}
\BIBentryALTinterwordspacing
Mozilla., ``{Seccomp filter in Android O},'' July 2016. [Online]. Available:
  \url{https://wiki.mozilla.org/Security/Sandbox/Seccomp}
\BIBentrySTDinterwordspacing

\bibitem{Mueller2020}
J.~M{\"u}ller, F.~Ising, C.~Mainka, V.~Mladenov, S.~Schinzel, and J.~Schwenk,
  ``Office document security and privacy,'' in \emph{{WOOT}}, 2020.

\bibitem{mulliner2015breaking}
C.~Mulliner and M.~Neugschwandtner, ``Breaking payloads with runtime code
  stripping and image freezing,'' \emph{BlackHat USA}, August 2015.

\bibitem{Narayan2020retrofitting}
S.~Narayan, C.~Disselkoen, T.~Garfinkel, N.~Froyd, E.~Rahm, S.~Lerner,
  H.~Shacham, and D.~Stefan, ``{Retrofitting Fine Grain Isolation in the
  Firefox Renderer},'' in \emph{{USENIX Security Symposium}}, 2020.

\bibitem{Nergal2001ret2libc}
Nergal, ``{The advanced return-into-lib(c) explits: PaX case study},'' 2001.

\bibitem{Prevelakis2001sandboxing}
V.~Prevelakis and D.~Spinellis, ``Sandboxing applications,'' in \emph{USENIX
  ATC}, 2001.

\bibitem{qian2019razor}
C.~Qian, H.~Hu, M.~Alharthi, P.~Ho~Chung, T.~Kim, and W.~Lee, ``{RAZOR}: A
  framework for post-deployment software debloating,'' in \emph{{USENIX}
  Security Symposium}, 2019.

\bibitem{Quach2017}
A.~Quach, R.~Erinfolami, D.~Demicco, and A.~Prakash, ``{A multi-OS cross-layer
  study of bloating in user programs, kernel and managed execution
  environments},'' in \emph{FEAST}, 2017.

\bibitem{quach2018debloating}
A.~Quach, A.~Prakash, and L.~Yan, ``Debloating software through piece-wise
  compilation and loading,'' in \emph{{USENIX Security Symposium}}, 2018.

\bibitem{Quynh2014capstone}
N.~A. Quynh, ``Capstone: Next-gen disassembly framework,'' \emph{Black Hat
  USA}, 2014.

\bibitem{Rajagopalan2005}
M.~Rajagopalan, M.~Hiltunen, T.~Jim, and R.~Schlichting, ``{Authenticated
  system calls},'' in \emph{DSN}, 2005.

\bibitem{Reis2009}
C.~Reis and S.~D. Gribble, ``{Isolating web programs in modern browser
  architectures},'' in \emph{EuroSys}, 2009.

\bibitem{Reis2019SiteIsolation}
C.~Reis, A.~Moshchuk, and N.~Oskov, ``{Site Isolation: Process Separation for
  Web Sites within the Browser},'' in \emph{USENIX Security Symposium}, 2019.

\bibitem{Schuster2015COOP}
F.~Schuster, T.~Tendyck, C.~Liebchen, L.~Davi, A.~Sadeghi, and T.~Holz,
  ``Counterfeit object-oriented programming: On the difficulty of preventing
  code reuse attacks in {C++} applications,'' in \emph{{S\&P}}, 2015.

\bibitem{Schwarz2019ZL}
M.~Schwarz, M.~Lipp, D.~Moghimi, J.~Van~Bulck, J.~Stecklina, T.~Prescher, and
  D.~Gruss, ``{ZombieLoad: Cross-Privilege-Boundary Data Sampling},'' in
  \emph{CCS}, 2019.

\bibitem{Schwarz2019SGXMalware}
M.~Schwarz, S.~Weiser, and D.~Gruss, ``{Practical Enclave Malware with Intel
  SGX},'' in \emph{DIMVA}, 2019.

\bibitem{Shacham2007}
H.~Shacham, ``{The geometry of innocent flesh on the bone: Return-into-libc
  without function calls (on the x86)},'' in \emph{CCS}, 2007.

\bibitem{Silvanovich2020}
N.~Silvanovich, ``{Exploiting Android Messengers with WebRTC: Part 1},'' 2020.

\bibitem{song2018fine}
L.~Song and X.~Xing, ``Fine-grained library customization,'' in \emph{SALAD},
  2018.

\bibitem{sqlite2020testing}
SQLite, ``{How SQLite Is Tested},'' 2020.

\bibitem{Steensgaard1996points}
B.~Steensgaard, ``Points-to analysis in almost linear time,'' in \emph{POPL},
  1996.

\bibitem{Sylvain2015}
\BIBentryALTinterwordspacing
N.~Sylvain, ``{A new approach to browser security: the Google Chrome
  Sandbox},'' October 2008. [Online]. Available:
  \url{https://blog.chromium.org/2008/10/new-approach-to-browser-security-google.html}
\BIBentrySTDinterwordspacing

\bibitem{Szekeres2013sok}
L.~Szekeres, M.~Payer, T.~Wei, and D.~Song, ``{SoK: Eternal War in Memory},''
  in \emph{S\&P}, 2013.

\bibitem{tarjan1972depth}
R.~Tarjan, ``Depth-first search and linear graph algorithms,'' \emph{SIAM
  journal on computing}, June 1972.

\bibitem{Mitre}
\BIBentryALTinterwordspacing
{The MITRE Corporation}, ``Common vulnerabilities and exposures.'' [Online].
  Available: \url{http://cve.mitre.org/}
\BIBentrySTDinterwordspacing

\bibitem{Tizen2014}
\BIBentryALTinterwordspacing
Tizen, ``{Security:Seccomp},'' 2018. [Online]. Available:
  \url{https://wiki.tizen.org/Security:Seccomp}
\BIBentrySTDinterwordspacing

\bibitem{VanSchaik2019RIDL}
S.~van Schaik, A.~Milburn, S.~Österlund, P.~Frigo, G.~Maisuradze, K.~Razavi,
  H.~Bos, and C.~Giuffrida, ``{RIDL: Rogue In-flight Data Load},'' in
  \emph{{S\&P}}, 2019.

\bibitem{wagner2000}
D.~Wagner and R.~Dean, ``Intrusion detection via static analysis,'' in
  \emph{S\&P}, 2000.

\bibitem{Wang2017angr}
F.~Wang and Y.~Shoshitaishvili, ``{Angr - The Next Generation of Binary
  Analysis},'' in \emph{2017 IEEE Cybersecurity Development (SecDev)}, 2017.

\bibitem{Weiser2019SGXJail}
S.~Weiser, L.~Mayr, M.~Schwarz, and D.~Gruss, ``{SGXJail}: Defeating enclave
  malware via confinement,'' in \emph{RAID}, 2019.

\bibitem{Firefox2019fission}
\BIBentryALTinterwordspacing
M.~Wiki, ``{Project Fission},'' 2019. [Online]. Available:
  \url{https://wiki.mozilla.org/Project_Fission}
\BIBentrySTDinterwordspacing

\bibitem{Firefox2019sandbox}
\BIBentryALTinterwordspacing
------, ``{Security/Sandbox},'' December 2019. [Online]. Available:
  \url{https://wiki.mozilla.org/Security/Sandbox}
\BIBentrySTDinterwordspacing

\bibitem{Yarom2014Flush}
Y.~Yarom and K.~Falkner, ``{Flush+Reload: a High Resolution, Low Noise, L3
  Cache Side-Channel Attack},'' in \emph{USENIX Security Symposium}, 2014.

\end{thebibliography}
